\documentclass[10pt]{iopart}

\usepackage{graphicx}
\expandafter\let\csname equation*\endcsname\relax
\expandafter\let\csname endequation*\endcsname\relax
\usepackage{amsmath}
\usepackage{amsfonts}
\usepackage{amssymb}
\usepackage{tikz}
\usepackage{hyperref}
\usepackage[nosort]{cite}

\newcommand*\circled[1]{\tikz[baseline=(char.base)]{
    \node[shape=circle,draw,inner sep=0.1pt] (char) {#1};}}

\usepackage{booktabs}
    
\begin{document}

\title[Characterisation of single microdischarges during PEO of Al and Ti]{Characterisation of single microdischarges during plasma electrolytic oxidation of aluminium and titanium}

\author{Jan-Luca Gembus$^{1}$, Vera Bracht$^{1}$, Florens Grimm$^{1}$, Nikita Bibinov$^{1}$, Lars Schücke$^{1}$, Peter Awakowicz$^{1}$, and Andrew R. Gibson$^{1, 2}$}

\address{
$^{1}$Chair of Applied Electrodynamics and Plasma Technology, Faculty of Electrical Engineering and Information Technology, Ruhr University Bochum, Germany \\
$^{2}$York Plasma Institute, School of Physics, Engineering and Technology, University of York, United Kingdom}
\ead{gembus@aept.rub.de}
\vspace{10pt}
\begin{indented}
\item April 2025
\end{indented}

\begin{abstract}
Plasma electrolytic oxidation (PEO) is a technique used to create oxide-ceramic coatings on lightweight metals, such as aluminium, magnesium, and titanium. PEO is known for producing coatings with high corrosion resistance and strong adhesion to the substrate. The process involves generating short-lived microdischarges on the material surface through anodic dielectric breakdown in a conductive aqueous solution.
To investigate single microdischarges during PEO, a single microdischarge setup was developed, where the active anode surface is reduced to the tip of a wire with a diameter of 1\,mm. In this work the focus is on the effect of electrolyte concentration, anode material, and electrical parameters on the microdischarges. The electrolyte is composed of distilled water with varying concentrations of potassium hydroxide (0.5\,- 4\,g/l). High-speed optical measurements are conducted to gain insights into the formation and temporal evolution of individual microdischarges and the induced gas bubble formation. Optical emission spectroscopy is used to estimate surface and electron temperatures by fitting Bremsstrahlung and Planck's law to the continuum spectrum of the microdischarges. To evaluate the impact of the microdischarges on coating morphology, the resulting oxide layers on the metal tips are analysed using scanning electron microscopy.
The study demonstrates that microdischarge behaviour is significantly influenced by the substrate material, treatment time, and electrolyte concentration, all of which impact the coating morphology. Under the conditions studied in this work, aluminium exhibits longer microdischarge and bubble lifetimes, with fewer cracks on the top layer of the coating, whereas titanium showed faster, shorter-lived bubbles due to more rapid microdischarge events.
\end{abstract}

%
%
%
%
\ioptwocol
%

\section{Introduction: plasma electrolytic oxidation}
Light metals like aluminium, titanium, magnesium, and their alloys are extensively used in transport and medical applications \cite{aliofkhazraeiReviewPlasmaElectrolytic2021, light_alloys_prefare}.
With the densities of aluminium (2.7\,g/cm$^3$) and magnesium (1.7\,g/cm$^3$) being less than one-third that of iron (8.9\,g/cm$^3$), they are suitable for weight reduction in car manufacturing, which has the potential to lead to reduced emissions from vehicles and increased driving range. However, due to higher costs and more complex processing, titanium is primarily used in medical implants, where its non-toxic properties and biocompatibility make it an ideal choice \cite{Light_metals_car_2021,civantos_ti_2017,alipal_2021}.
All of these metals have a high affinity for oxygen, causing the generation of a thin passivation layer on their surface, which protects them from corrosion and wear in normal environments \cite{light_alloys_prefare,ferraris_surface_2021, polmear_light_2017}.
The long-term durability of this layer is limited, especially in harsh environments, such as car engines or medical implants \cite{Light_metals_car_2021,costa_synthesis_2020}. It can be easily damaged, resulting in material degradation. Additional protection is necessary to increase the lifespan of the components and extend their usage to further applications. Common surface treatments such as anodising, hard anodising, or plasma spraying can enhance the durability of the substrate, but are often limited by alloy compatibility and the potential for defects and cracks \cite{PEO_book_2010_chapter05,keronite_2018}. \\

Plasma electrolytic oxidation (PEO) is a surface passivation process that can be used to create an oxide coating providing reliable corrosion and wear protection \cite{ProductionAntiCorrosionCoatings2014}. It can also enhance the biocompatibility of titanium or magnesium implants by incorporating elements such as silicon or phosphorus into the passivation layer, or through surface modifications that improve corrosion resistance \cite{mohedano_bioactive_2013,husak_bioactivity_2021}.
In contrast to anodizing, the PEO process involves higher voltages ($\approx$\,250\,-\,750\,V) and typically alkaline electrolytes instead of acids, which can make it more environmentally friendly \cite{ProductionAntiCorrosionCoatings2014,huangPlasmaElectrolyticOxidation2019, sikdarPlasmaElectrolyticOxidation2021, husseinInvestigationCeramicCoating2013, clyne_2018_review}.
In PEO the substrate is submerged in an electrolytic bath and acts as the anode. The process starts by anodic oxidation, generating an initial passivation layer on the substrate.
Once the substrate is fully covered and the layer reaches sufficient thickness, dielectric breakdown occurs, leading to the formation of short-living microdischarges on the passivation layer \cite{husseinInvestigationCeramicCoating2013}. These are accompanied by gas production in the form of growing bubbles, which contain oxygen, water vapour, and hydrogen \cite{clyne_2015,coatings_PEO_simchen}.
Initially, only small pores are created in the passivation layer, with a characteristic melting and solidification of the substrate material. As the coating thickness increases, the number of microdischarges decreases over time, leading to a rougher outer layer \cite{hermanns_-situ_2020,vargasMorphologicalAnalysisPlasma2023,lucasPlasmaElectrolyticOxidation2024}. A further increase of the voltage causes larger, arc-like discharges, which potentially damage the coating \cite{clyne_2018_review, hermanns_-situ_2020}. The different types of discharges occurring during the process can be further classified. One concept to do so as proposed by Hussein~\textit{et al.}~\cite{husseinSpectroscopicStudyElectrolytic2010}, who identified three types of discharges involved in the coating growth. Type B discharges, occurring at the metal-oxide interface, type C, occurring at the oxide-electrolyte interface within the upper coating and type A, occurring at the top layer of the coating.
Typically, microdischarge formation creates a coating with characteristic cracks and holes, making it less uniform compared to anodised coatings \cite{sikdarPlasmaElectrolyticOxidation2021}. 
On the other hand, PEO processes produce coatings with a thickness up to hundreds of \textmu m and they can be uniformly distributed, even on edges, avoiding common issues like cracking, which can be seen in hard anodising processes \cite{keronite_2018, sikdarPlasmaElectrolyticOxidation2021}. \\

In general, the coating produced in a PEO process is influenced by a variety of parameters like electrical parameters, substrate composition, deposition time, as well as electrolyte concentration and composition \cite{coatings_PEO_simchen}.
Because the process is driven by short lived microdischarges, the coating is expected to be affected by the number, duration, and intensity of these microdischarges \cite{husseinInvestigationCeramicCoating2013}. However, their transient nature, their dependency on multiple parameters, and their occurrence in liquid solutions lead to a highly complex system with limited diagnostic methods available to characterise them \cite{clyne_2018_review}. While most research has focused on larger scaled coating setups on specific substrates, for example \cite{sikdarPlasmaElectrolyticOxidation2021, clyne_2018_review, coatings_PEO_simchen, hermanns_-situ_2020, lucasCharacterizationOxideCoating2022, brachtModificationsElectrolyticAluminum2021}, studies on individual microdischarges remain limited, such as \cite{clyne_2015,clyne2016}.
This study focuses on investigating the effects of electrolyte concentration and deposition time in greater detail, using aluminium and titanium anodes. Gaining a deeper understanding of the transient microdischarges occurring during the process may provide valuable insights for controlling the coating properties.

\section{Experimental setup}
\begin{figure}
    \centering
    \includegraphics[width=8cm]{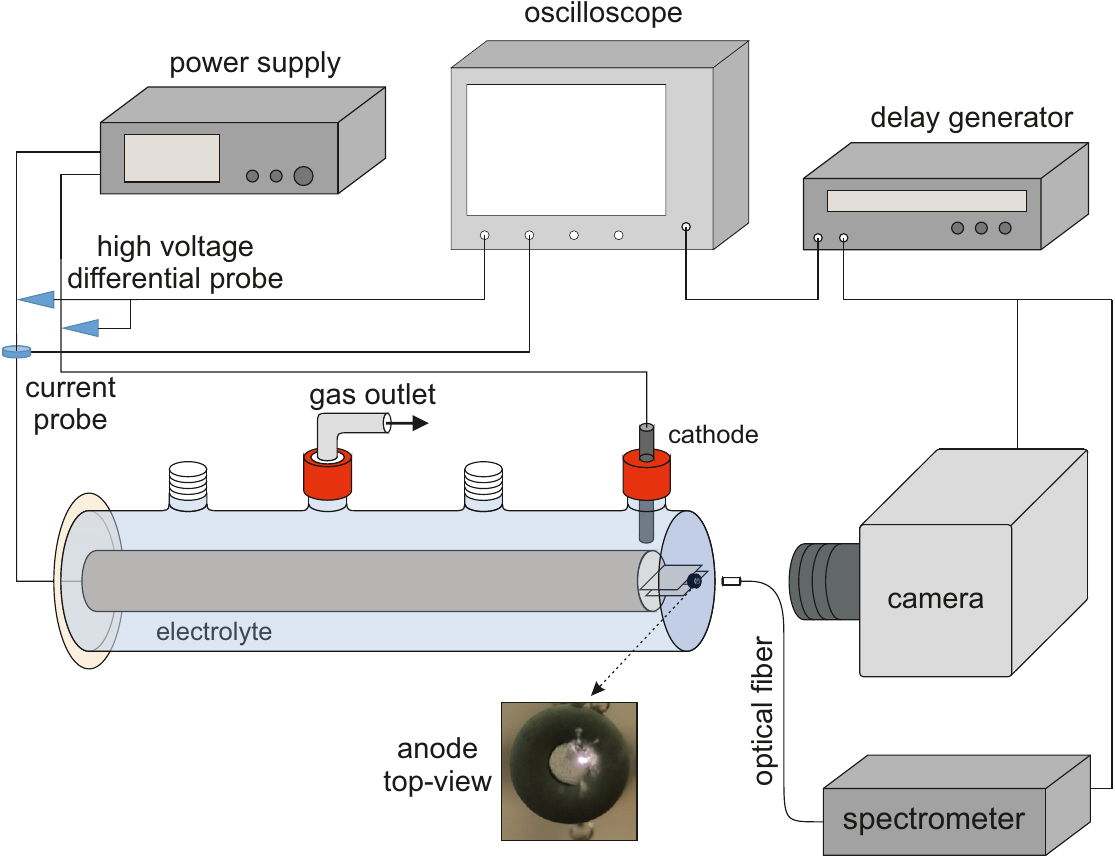}
    \caption{Schematic of a single microdischarge (SMD) setup. It allows the observation of single microdischarges during a PEO process by reducing the anode to the tip of a wire with a diameter of 1\,mm. The anode is immersed in an electrolytic cell. A quartz glass window in line-of-sight to the substrate tip allows an investigation of single discharges.} 
    \label{SMD_setup01}
\end{figure}

An experimental system specialised for the study of single microdischarges (SMDs) during PEO has been developed. This system was originally described in detail in the work of Bracht \cite{Bracht_diss}. The reduction of the anode/substrate to the tip of a wire with a diameter of 1\,mm ensures ignitions of mainly single discharges on the wire tip, each with a lifetime of few to hundreds of \textmu s. An illustration of the setup is shown in figure~\ref{SMD_setup01}. It is based on a quartz-glass tube, which allows optical measurements in line-of-sight of the substrate surface. On one side of the glass tube a KF40 glass flange is attached, where a polyether ether ketone (PEEK) holder is mounted and sealed with an O-ring. The holder is used to fix the substrate wire inside the glass tube, with a gap of approximately 1\,cm between the tip of the substrate and the quartz window. This allows microdischarges to be investigated by optical measurements in line-of-sight to the substrate. The tube has four outlets at the top, where one is used as gas outlet and another one to connect the counter electrode. The last two can be used to insert additional diagnostic tools, such as a Pt100 temperature sensor for measuring liquid temperature.
In this work, an aluminium ($\mathrm{Puratronic}^{\circled{\tiny{R}}}$, 99.9995\,\%) or titanium (Alfa Aesar, 99.99\,\%) wire is used as anode substrate material, while the cathode is made of stainless steel. The substrate wire is insulated with a shrinking tube and an additional O-ring at the tip. The heat resistant fluorine rubber ($\mathrm{Viton}^{\circled{\tiny{R}}}$) O-ring is needed as protection due to locally high temperatures up to few 1000\,K at the substrate surface, which could potentially melt the insulation.
The liquid electrolyte in this study is a mixture of distilled water ($\leq$\,700\,\textmu S/cm) and a varying amount of potassium hydroxide (0.5\,-\,4\,g/l of KOH, $\geq$\,85\,\%). The resulting conductivity ranges from 2 to 16.7\,mS/cm, with a pH value of approximately 12, depending on the concentration used, as shown in table~\ref{tab:Electrolyte}. The electrolytic cell contains around 180\,ml of the liquid. The temperature variation of the liquid caused by short and local hot discharges remains minimal with approximately $\Delta$T\,=\,2.5\,K over a 10 minute treatment, which corresponds to the process time in this study. \\

The power supply (2260B-800-1, Keithley) is operated in a galvanostatic DC-mode with a fixed current density of 1.27\,A/cm$^{2}$ at the anode surface and a maximum voltage output of 800\,V, ensuring the process is not limited by the voltage output. Bracht's work \cite{Bracht_diss} investigated the same current density of 1.27\,A/cm$^{2}$ as well as a lower current density of 0.64\,A/cm$^{2}$ for an aluminium substrate. Despite a better formation of single microdischarges for aluminium at lower current density, the higher current density was selected to ensure microdischarge ignition with a titanium substrate, as no microdischarges were ignited at lower current densities on titanium. 
Electrical parameters are monitored with a high voltage differential probe (HVD3102A, Teledyne LeCroy) between anode and cathode as well as a current probe (CP030A, Teledyne LeCroy) at the anode. Both probes are connected to an oscilloscope (WaveRunner 8254, Teledyne LeCroy) with a maximum sample rate of 20 GS/s.
Triggering on the microdischarge current enables time-synchronized measurements with different diagnostic tools such as spectrometers and cameras. Additionally, a delay generator (DG535, Stanford Research Systems) is used to produce precise trigger pulses and adjust the timing to analyse different stages of a microdischarge lifecycle. \\

Before each PEO treatment the substrate tip is polished with a standard commercial polishing file to remove any residue of previous treatments and to ensure a flat surface at the start of the process. The considered treatment time of 10 minutes begins with the ignition of the first microdischarges. That means, for an aluminium substrate the process starts almost immediately after the power is applied, whereas for a titanium substrate the process requires some time to start, which can be up to 5 minutes. This is caused by different formation mechanisms of the initial oxide layer on the substrate surface, influenced by variations in electrical conductivities of the substrate and the resulting oxide layer. 

\begin{table}[]
\centering
\begin{tabular}{@{}cccc@{}}
\toprule
\makebox[6em]{\shortstack{KOH ($\geq$\,85\,$\%$) \\ (g/l)}} & \makebox[4em]{\shortstack{Molarity \\ (M)}} & \makebox[5em]{\shortstack{Conductivity \\ (S/m)}} &   \makebox[3em]{\shortstack{\vspace{5pt} pH}} \\ \midrule
    0.5     & 0.0089   & 0.23  & 12.1 \\ 
    1       & 0.0178   & 0.43  & 12.3 \\
    2       & 0.0356   & 0.82  & 12.6 \\
    3       & 0.0535   & 1.24  & 12.8 \\
    4       & 0.0712   & 1.67  & 12.9 \\
    \bottomrule
    \end{tabular}
    \caption{Measured electrical conductivity and pH value for the used electrolyte concentration between 0.5\,g/l and 4\,g/l KOH in distilled water.}
    \label{tab:Electrolyte}
\end{table}

\section{Diagnostic methods}
Different diagnostic tools are applied to observe the substrate tip and investigate individual microdischarges and bubble dynamics during the PEO process (sections \ref{High_speed_imaging}, \ref{continuum_measurements}). In addition, a post treatment analysis is performed with a scanning electron microscope (SEM) in combination with an energy-dispersive X-ray spectrometer (EDX) to study the morphology and composition of the created coating (section~\ref{SEM}).

\subsection{High speed imaging}\label{High_speed_imaging}
A high-speed camera (VEO 410L IMP, Vision Research) operating at $150\,000$~fps is combined with backlighting. This enables the observation of individual bubble and microdischarge formation during the PEO process. The high frame rate is achieved by using only $128\,\times\,128$ pixels of the camera sensor, resulting in a time interval of 6.66\,\textmu s between images. As backlight, an ultra-high-pressure lamp (UHP 100\,W/120\,W 1.0, Phillips) is focused on the substrate tip to observe bubble formation. A disadvantage of the backlighting is the reflection of the light source on the bubble surface, which can sometimes be challenging to differentiate from the bright microdischarge at the bubble's centre. The camera is triggered on the microdischarge current, allowing for the synchronous measurement of the electrical parameters with the bubble radius. However, limiting the resolution to 128 pixels introduces some inaccuracy in determining the bubble radius, allowing for a minimum measurable bubble radius of approximately 8\,\textmu m. The pixel size of the image is calculated based on the known diameter of the substrate surface, which is 1\,mm. Continuous image recording of the camera allows capturing images at all stages of the bubble lifetime. \\

Furthermore, the images are used to investigate the gas pressure development inside a single bubble. For this, the Rayleigh-Plesset equation (see equation~\eqref{equ:rayleigh}) is used to model the spherical cavitation in an incompressible liquid. By combining this model with the time-dependent bubble radius measurements obtained through high-speed imaging, it is possible to estimate the pressure inside a bubble during the PEO process. However, several assumptions are made to apply this equation, as demonstrated in a similar study by Troughton~\textit{et al.}~\cite{clyne_2015}. First, bubble formation at an electrode interface leads to a hemispherical bubble. Since no model currently exists to describe dynamics for a hemispherical bubble, the bubble is assumed to be spherical, with expansion and collapse slower than the speed of sound \cite{hamdanDynamicsBubblesCreated2013}. Furthermore, the liquid is treated as an incompressible and homogeneous Newtonian fluid with constant properties, while external forces like gravity and centrifugal effects are ignored. Lastly, it should be noted that gradients of temperature and pressure are also neglected \cite{rayleigh_Tey_2020}.  The Rayleigh-Plesset equation is used in the following form:

\begin{equation}\label{equ:rayleigh}
    R\frac{\mathrm{d}^{2}R}{\mathrm{d}t^{2}}+\frac{3}{2}\left(\frac{\mathrm{d}R}{\mathrm{d}t}\right)^{2}+\frac{4 \eta}{\rho R}\frac{\mathrm{d}R}{\mathrm{d}t}+\frac{2 \sigma_{\mathrm{s}}}{\rho R}=\frac{1}{\rho}\left(P_{\mathrm{b}}-P_{\infty}\right)
\end{equation}
Here, $R$ is the time-dependent radius of the bubble, $\eta$ the dynamic viscosity of the liquid, $\sigma_{\mathrm{s}}$ the surface tension coefficient between the liquid and gas vapour interface, $\rho$ the density of the liquid, $P_{\mathrm{b}}$ the pressure of the bubble and lastly, $P_{\infty}$ the pressure at a large distance of the bubble \cite{rayleigh_equation}.

\subsection{Measurements of PEO emission spectrum}\label{continuum_measurements}
A compact low resolution spectrometer (QE65000, Ocean Optics) is used to measure the emission spectrum during the PEO process. It is relative calibrated based on the study by Gr\"oger~\textit{et al.}\cite{grogerCharacterizationTransientSpark2020}.
It has a cooled back-thinned detector (S7031-1006, Hamamatsu) with sensitivity in the range of 200 to 1100\,nm, though it is practically usable up to 900\,nm due to the grating used. The spectral resolution is approximately 1.3\,nm and it is connected to a 600\,\textmu m diameter quartz fiber (FOA FDP600660710 UVM, LLA Instruments). The integration time was set to 500\,ms for Al, with measurements averaged over 20 spectra. Due to lower radiation intensity with a Ti substrate, the integration time was increased to 5000\,ms, with an average of 5 spectra. Measurements are performed every minute during the 10 minute PEO process, with each measurement including hundreds of discharges. To increase the measured signal, a collimator is connected to the fiber and placed in line-of-sight to the substrate surface. The emission spectra of PEO consists of atomic lines, molecular bands, and a continuum. Fitting a theoretically calculated continuum spectrum to the measured spectrum allows for estimation of the surface temperature and electron temperature, as described in the following. In this work, we interpret the surface temperature as a lower boundary for the gas temperature. \\

Continuum radiation can result from free electron interactions with neutrals and ions, which can be separated into free-free and free-bound interactions \cite{Bilek_2021}. Free-free interactions occur during electron interactions with neutrals or ions and correspond to Bremsstrahlung radiation, while free-bound interactions occur through electron-ion recombination and result in recombination radiation. In addition, thermal radiation from the substrate is observed. It is caused by local microdischarges that heat the substrate, leading to melting and even boiling of the substrate material. By approximating the substrate material as a black body radiator (surface emission coefficient $\epsilon\,=\,1$), its thermal radiation can be described by Planck's law \cite{Bilek_2021,intro_plasma_spectroscopie_2009}. \\

First, only free-free and free-bound particle interactions are taken into account. The resulting total emission coefficient, $\epsilon_{\mathrm{cont}}$, is the sum of the individual emission coefficients, as illustrated by the following equation \cite{Bilek_2021,KTALBurm_2004}:
\begin{equation}    
    \epsilon_{\mathrm{cont}}=\epsilon_{\mathrm{en}}^{\mathrm{ff}}+\epsilon_{\mathrm{ei}}^{\mathrm{ff}}+\epsilon_{\mathrm{ei}}^{\mathrm{fb}}    
\end{equation}
where free-free electron-neutral interactions are described by $\epsilon_{\mathrm{en}}^{\mathrm{ff}}$, free-free electron-ion interactions by $\epsilon_{\mathrm{ei}}^{\mathrm{ff}}$ and free-bound electron-ion recombinations by $\epsilon_{\mathrm{ei}}^{\mathrm{fb}}$. \\

The PEO process typically involves significantly higher neutral particle density than ion density ($n_\mathrm{0}\gg n_\mathrm{i}\approx n_\mathrm{e} (\mathrm{max}) \approx 10^{17}\,\mathrm{cm^{-3}}$) 
resulting in most of the emission being driven by electron-neutral interactions \cite{clyne_2015, Bracht_diss}. Consequently, the two coefficients involving ion interactions are neglected ($\epsilon_{\mathrm{cont}}\approx\epsilon^{\mathrm{ff}}_{\mathrm{en}}$). The emission coefficient $\epsilon^{\mathrm{ff}}_{\mathrm{en}}$ is given by the following equation \cite{Bilek_2021}:
\begin{equation}\label{equ:cont_ff01}
    \epsilon^{\mathrm{ff}}_{\mathrm{en}}(\lambda)=\sqrt{\frac{2}{m_\mathrm{e}}}\frac{n_\mathrm{e} n_\mathrm{n}}{\lambda^2}\frac{hc}{4\pi}\int_{\frac{hc}{\lambda}}^{\infty}\nu\frac{\mathrm{d}\sigma_{\mathrm{el}}}{\mathrm{d}\nu} E^{1/2} f(E) \mathrm{d}E
\end{equation}
where $m_\mathrm{e}$ is the electron mass, $n_\mathrm{e}$ and $n_\mathrm{n}$ are the densities of electrons and neutrals respectively, $h$ is the Planck constant, $c$ is the speed of light, $\sigma_{el}$ is the cross section for elastic scattering and $f(E)$ is the normalised electron energy distribution function (EEDF). Additionally, the emitted photon is characterised by its wavelength $\lambda$ and frequency $\nu$. \\

To simplify equation~\eqref{equ:cont_ff01} and derive an expression in dependency of the electron temperature, several assumptions are necessary \cite{Bilek_2021}. The polarization effects of neutrals are neglected, and a weak energy dependency of the cross-section $\sigma_{el}$ is assumed. Lastly, $f(E)$ is assumed to be a Maxwellian distribution function, allowing equation~\eqref{equ:cont_ff01} to be rewritten as equation~\eqref{equ:cont_ff02} in units of W/(m$^4$sr$^1$). \\ 
\begin{equation} \label{equ:cont_ff02}
\begin{split}
\epsilon_{\mathrm{ff}}^{\mathrm{en}}(\lambda, T_\mathrm{e}) = \,& C_{1}^{''}\frac{n_{\mathrm{e}}n_{\mathrm{n}}}{\lambda^{2}}\left(k_\mathrm{{B}}T_{\mathrm{e}}\right)^{\frac{3}{2}} \langle Q^{\mathrm{m}}_{\mathrm{en}}(T_{\mathrm{e}}) \rangle \\
 & \times\left[\left(1+\frac{hc}{\lambda k_{\mathrm{B}}T_{\mathrm{e}}}\right)^{2}+1\right]\mathrm{exp}\left(\frac{-hc}{\lambda k_\mathrm{B}T_\mathrm{e}}\right)
\end{split}
\end{equation}
Here $C_1^{''}$\,=\,2.0\,J$^{-1/2}$m$^2$s$^{-1}$sr$^{-1}$ is a proportionality constant, $k_\mathrm{B}$ is the Boltzmann constant, $T_\mathrm{e}$ is the electron temperature, and $\langle Q^{\mathrm{m}}_{\mathrm{en}}(T_\mathrm{e})\rangle$ is the cross section for the electron-water molecule momentum transfer at the temperature $T_\mathrm{e}$, averaged over the Maxwellian distribution. The corresponding cross section is taken from Yousfi~\textit{et al.} \cite{continuum_const01}. The measured emission intensity is proportional to the number of emitted photons with a proportionality factor $C_3$, divided by the photon energy \cite{Bilek_2021}. This is given with $I_{\mathrm{ff}}^{\mathrm{en}}(\lambda)=C_3\epsilon_{\mathrm{ff}}^{\mathrm{en}}(\lambda)/E_{\mathrm{photon}}$. Further simplifications are made by approximating the densities and cross-section to be nearly constant as the spectrum is averaged over hundreds of discharges per measurement. Furthermore, only the effects over a 10 minute treatment time are considered. That allows them to be combined into a single scaling factor $C_{1}^{'}$. This results in the final expression for the intensity of Bremsstrahlung: \\
\begin{equation} \label{equ:cont_ff03}
\begin{split}
I_{\mathrm{ff}}^\mathrm{{en}}(\lambda, T_\mathrm{e}) = \,& C_{1}^{'}\frac{\left(k_\mathrm{B}T_\mathrm{e}\right)^{\frac{3}{2}}}{\lambda\,h\,c} \left[\left(1+\frac{hc}{\lambda k_\mathrm{B}T_\mathrm{e}}\right)^{2}+1\right] \\
 & \times \mathrm{exp}\left(\frac{-hc}{\lambda k_\mathrm{B}T_\mathrm{e}}\right)
\end{split}
\end{equation}

As previously noted, thermal radiation from the anode surface also contributes to the continuum radiation. For simplicity, a black body radiator is assumed, with an intensity calculated using Planck's law divided by the energy $E=h\times c/\lambda$ \cite{hermanns_-situ_2020}. This is necessary to fit the calculated spectrum to the measured one and results in the following expression for the intensity of a black body radiator \cite{intro_plasma_spectroscopie_2009}: \\
\begin{equation}\label{equ:cont_BB}
    I_{\mathrm{BB}}(\lambda, T)=\frac{2\pi c\,C_{2}^{'}}{\lambda^{4}}\left(\mathrm{exp}\left(\frac{hc}{\lambda k_\mathrm{B}T}\right)-1\right)^{-1}
\end{equation}

In this equation $h$ is Planck's constant, $c$ is the speed of light, $\lambda$ the wavelength of the emitted light, $T$ the surface temperature, $k_\mathrm{B}$ the Boltzmann constant, and $C_2^{'}$ is a scaling factor. \\

To account for water absorption between the anode tip and the spectrometer, the intensity is adjusted using the Beer-Lambert law \cite{hermanns_-situ_2020,intro_plasma_spectroscopie_2009}. By applying this law and combining the intensities of Bremsstrahlung and black body radiation, the following final equation for the intensity of the continuum radiation \emph{I}\textsubscript{con} is obtained:
\begin{equation}\label{equ:cont_final}
\begin{split}
    I_{\mathrm{con}}(\lambda) = \,& (C_1 I_{\mathrm{ff}}^{\mathrm{en}}(T_\mathrm{e},\lambda)+C_2 I_{\mathrm{BB}}(T_\mathrm{s},\lambda)) \\
    &\times \mathrm{exp}{(-\alpha (\lambda)\,d)}
\end{split}
\end{equation}
Here $C_1$ and $C_2$ are scaling factors, $\alpha (\lambda)$ is the absorption coefficient of water \cite{water_abs_coeff_1973}, and $d$ is the distance between the anode and the quartz glass, which is approximately 1\,cm. By fitting the final equation~\eqref{equ:cont_final} to the measured spectrum, the electron and surface temperatures are obtained. As mentioned previously, here, the surface temperature is considered the lower boundary of the gas temperature at the surface of the substrate.

\subsection{Post treatment surface analysis}\label{SEM}
A scanning electron microscope (SEM, JSM6510, JEOL) in combination with an energy-dispersive X-ray spectrometer (EDX, XFlash Detector 410-M, Bruker) is used to characterise the morphology and composition of the generated coating. The SEM uses a high-energy electron beam to scan the surface of an object \cite{inkson_SEM_2016}. Non-conductive samples, like the oxide coatings grown on the substrate, interfere with the scanning process of the SEM due to charging effects caused by accumulation of electrons on the oxide surface. To prevent this, the samples are coated with a thin layer of gold, with an approximate thickness of 25\,nm, using a dedicated sputter coater (JFC-1200, JEOL). SEM and EDX measurements are performed on gold coated wire tips, which are cut off from the wire after a 10 minute PEO treatment. 

\section{Results and discussion}
This section is divided into four parts comparing the effect of electrolyte concentration and treatment time on an aluminium (Al) and a titanium (Ti) substrate during a PEO process. The first part investigates the current and voltage behaviour during the PEO process, followed by an analysis of bubble dynamics in relation to microdischarges. The third part focuses on determining surface- and electron temperatures using continuum radiation analysis. The last part explores the impact of PEO processing under different conditions on coating morphology.

\subsection{Current and voltage behaviour}\label{CV_section}

\begin{figure}
    \centering
    \includegraphics[width=8cm]{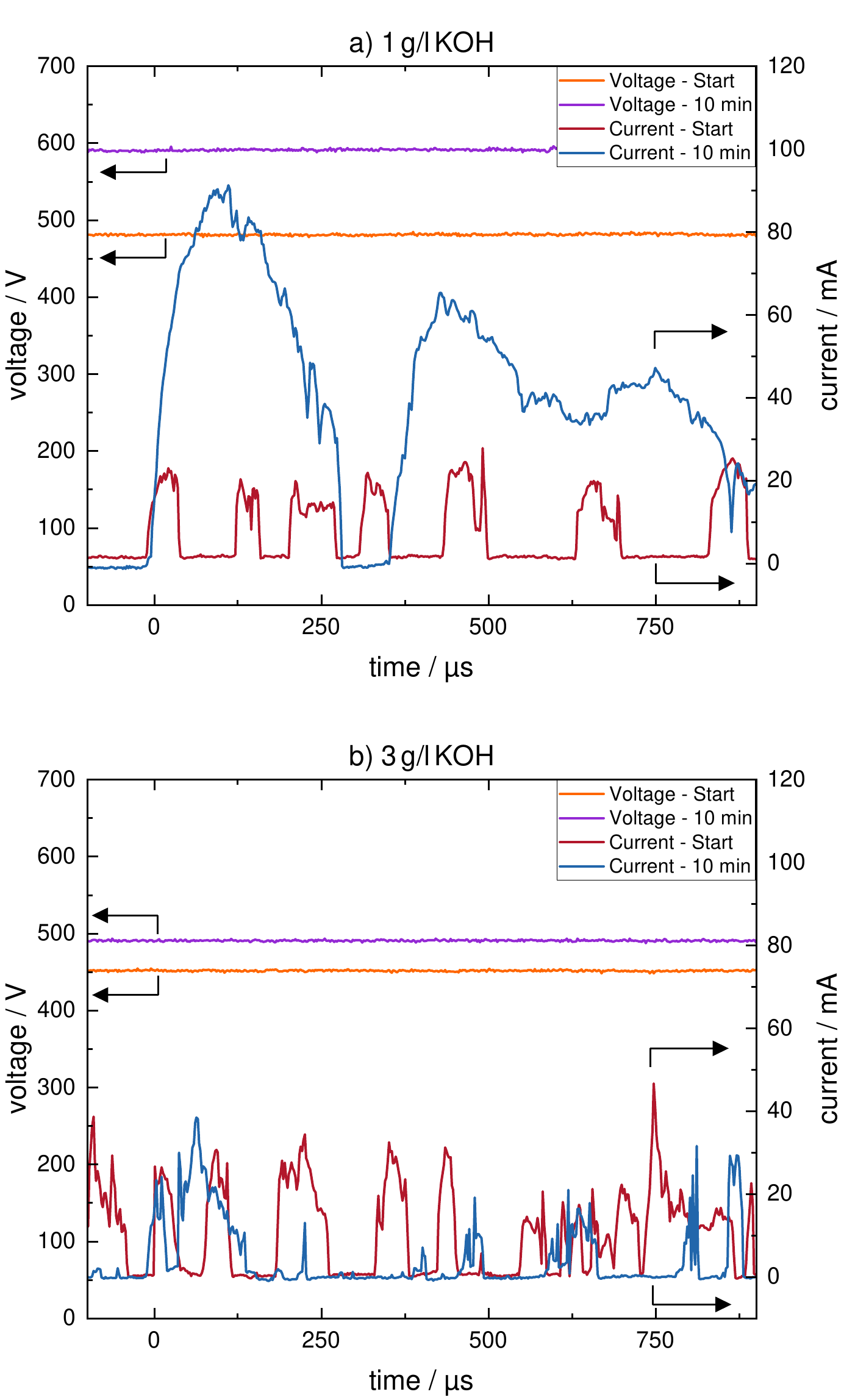}
    \caption{Time-resolved current-voltage measurements at the start and after 10 minute PEO treatment with an aluminium anode. The voltage is shown in orange (start), purple (10\,min) and the current in red (start), blue (10\,min). a) with 1\,g/l of KOH and b) with 3\,g/l of KOH in distilled water.} 
    \label{res:Al_CV_vert}
\end{figure}

\begin{figure}
    \centering
    \includegraphics[width=8cm]{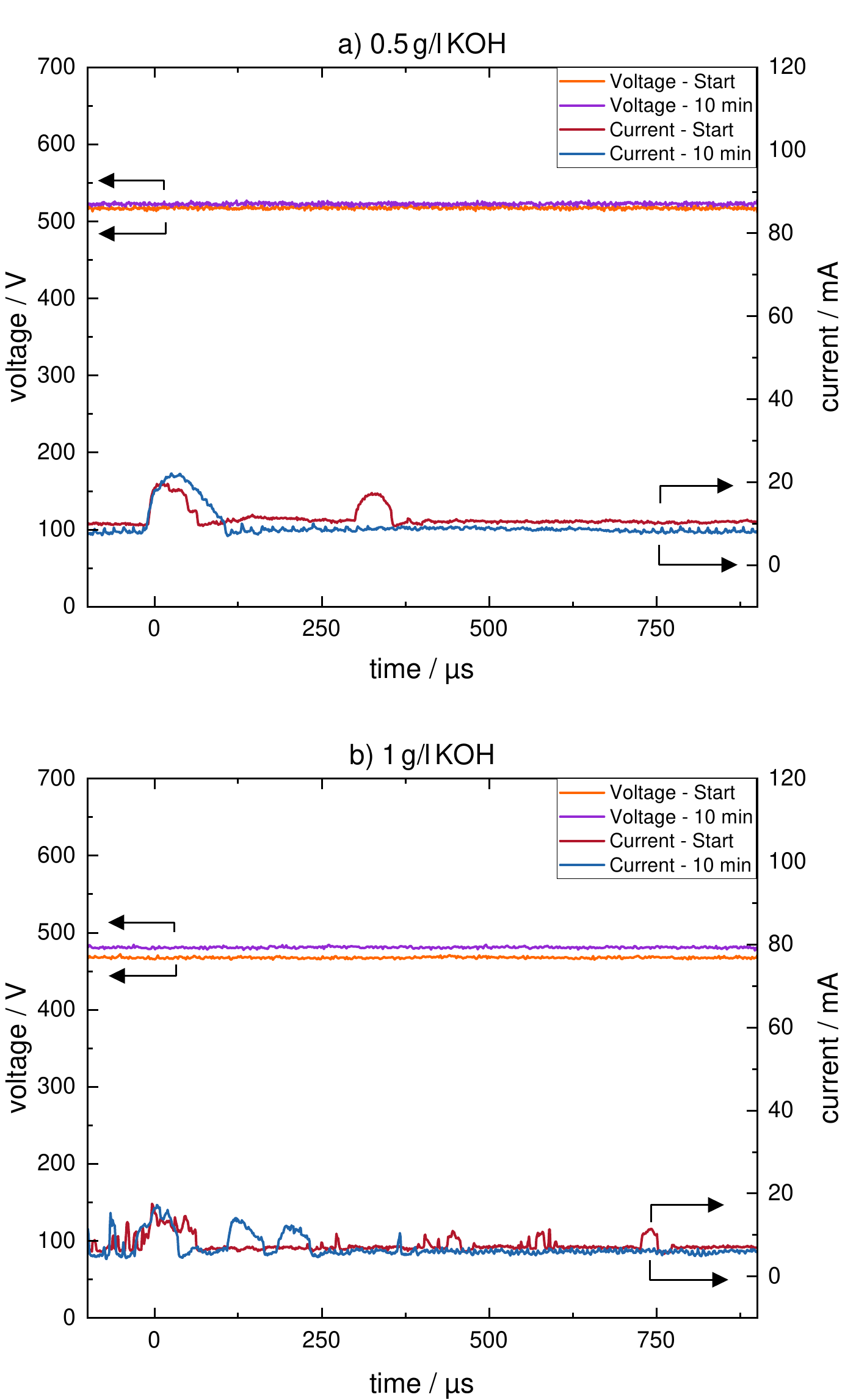}
    \caption{Time-resolved current-voltage measurements at the start and after 10\,min of PEO treatment with a titanium anode. The voltage is shown in orange (start), purple (10\,min) and the current in red (start), blue (10\,min) for a) 0.5\,g/l of KOH and b) 1\,g/l of KOH, in distilled water.} 
    \label{res:Ti_CV_vert}
\end{figure}

\begin{figure}
    \centering
    \includegraphics[width=8cm]{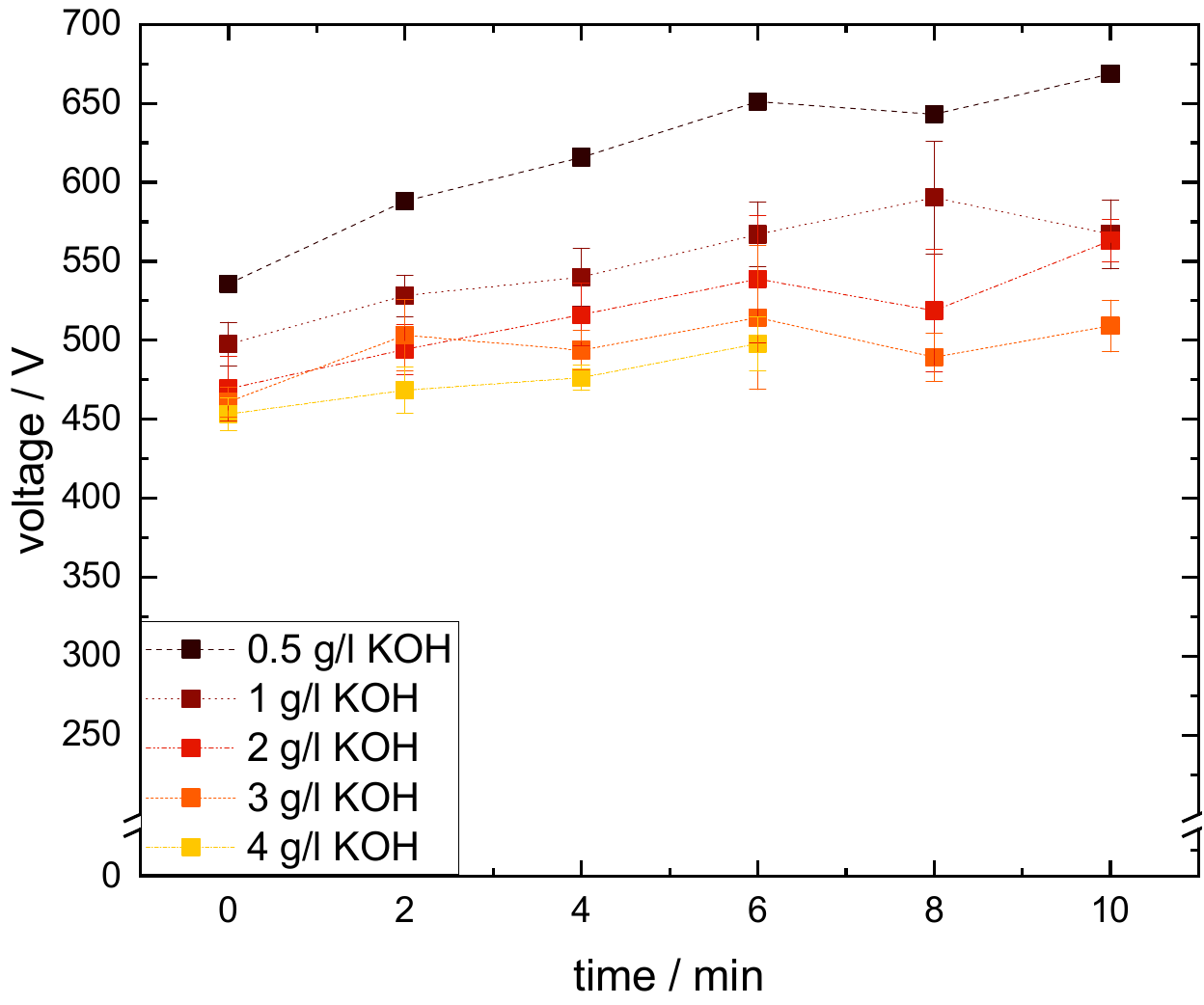}
    \caption{Voltage behaviour and corresponding standard deviation during a PEO treatment with an aluminium anode and different electrolyte (KOH) concentrations, while the current supply is fixed at 1.27\,A/cm$^2$. Higher voltages are measured for a lower KOH concentration and also for increasing treatment time.} 
    \label{res:Al_time02}
\end{figure}

\begin{figure}
    \centering
    \includegraphics[width=8cm]{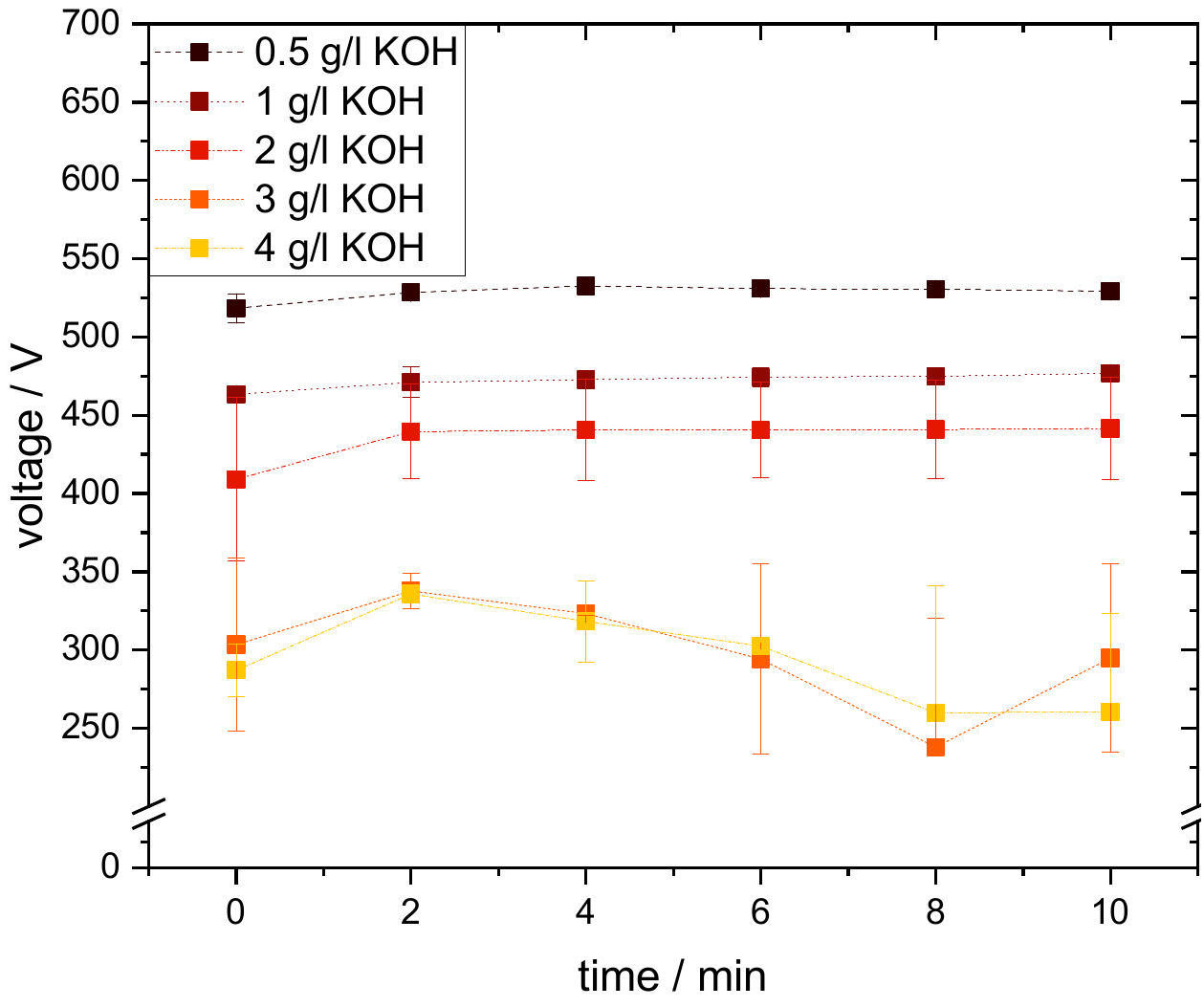}
    \caption{Voltage behaviour and corresponding standard deviation during a PEO treatment with a titanium anode and different electrolyte (KOH) concentrations, while the current supply is fixed at 1.27\,A/cm$^2$. Thereby, the voltage variation over time is relatively low for 1\,g/l and 2\,g/l of KOH. Concentrations above 2\,g/l of KOH are not reproducible and are accompanied with a high standard deviation.} 
    \label{res:Ti_time02}
\end{figure}

Figures~\ref{res:Al_CV_vert} and~\ref{res:Ti_CV_vert} show typical voltage and current behaviour for 900\,\textmu s of PEO treatment, after triggering on the microdischarge current at the start of the treatment (red, orange), and after another 10 minutes of treatment time (blue, purple). To highlight key results, only two electrolyte concentrations are shown for comparison: 1\,g/l and 3\,g/l for Al (figure~\ref{res:Al_CV_vert}), and 0.5\,g/l and 1\,g/l for Ti (figure~\ref{res:Ti_CV_vert}). 
Each current pulse represents an individual microdischarge, which is in agreement with Troughton~\textit{et al.} \cite{clyne_2015}. Multiple ignitions on the substrate lead to an overlapping total current, making individual current pulses indistinct and difficult to discern. The current of the Al and Ti substrate in figures~\ref{res:Al_CV_vert} and \ref{res:Ti_CV_vert} show that individual current pulses are successfully resolved with the setup used. It should be noted that while the microdischarges are highly transient and their exact forms are not very reproducible, the overall trends remain consistent. In contrast to the current, the voltage remains constant over these short periods of time ($\lessapprox$\,1\,ms), and does not change in response to the current pulses. It only increases over longer time periods with a constant current supply, which can be seen by the voltage over time measurements in figure~\ref{res:Al_time02} and \ref{res:Ti_time02}. \\

The behaviour of microdischarges during PEO treatment varies significantly depending on the electrolyte concentration and substrate material, which creates challenges in certain combinations. For example, with an Al substrate and a low electrolyte concentration of 0.5\,g/l of KOH  (not shown), multiple ignitions occur simultaneously. In contrast, with a Ti substrate, this behaviour is only observed at higher concentrations exceeding 1\,g/l of KOH. The reason for this is unclear, but it shows that a 1\,mm anode tip does not consistently produce single ignitions and is dependent on substrate materials and electrolyte concentrations.
Furthermore, concentrations exceeding 2\,g/l of KOH and with a Ti substrate result in no microdischarge ignition. Instead, rapid voltage fluctuations over 100\,V within milliseconds are observed, which is accompanied by a continuous current flow without any current pulses.
This behaviour can be broadly explained using the PEO model proposed by Hermanns~\cite{hermanns_diss}. In this model, the electrical circuit between the anode and cathode is represented by a combined capacitance of an oxide-layer and double-layer capacitance in parallel with an oxide resistance. The microdischarge ignition can be described by an additional plasma impedance parallel to it. The configuration is completed by an electrolyte resistance in series to account for the electrolyte-water solution.
Rapid voltage fluctuations indicate a continuous change in oxide resistance due to the generation and detachment of the coating. Furthermore, current flow may occur behind the fluorine rubber O-ring, both with and without discharge ignitions. A similar behaviour is observed for a process of 4\,g/l of KOH with an Al substrate, where the generation of single discharges on the tip of the substrate stops mostly after 6 minutes treatment time. Overall, single microdischarges over 10 minute treatment time are generated for 0.5\,-1\,g/l KOH with a Ti substrate and for 1\,-3\,g/l KOH with an Al substrate. In addition, concentration of 3\,g/l and 4\,g/l KOH show no PEO process with a Ti substrate.\\  

The probability of high current discharges increases over treatment time, which is induced by a growing passivation layer and can damage the O-ring at the tip of the anode. It was observed that a higher amount of KOH (e.g. 4\,g/l) leads to a higher probability of side ignitions behind the O-ring. This is assumed to be caused by high local temperatures close to the sealing, resulting in damage of the fluorine rubber O-ring and enabling ignitions in the resulting gap behind it. This behaviour is further promoted by the applied current density of 1.27\,A/cm$^2$. Reducing the current density to half of its initial value results in more distinct and reproducible current pulses with fewer current spikes. This is supported by Bracht's study \cite{Bracht_diss}, where the effect was less strong with half of the current density. Side ignition can likely be reduced by exchanging the fluorine rubber O-ring with one of a different material, which is planned for future experiments. \\

The lifetime of microdischarges can be estimated by investigating the current pulses (red and blue) in figures~\ref{res:Al_CV_vert} and \ref{res:Ti_CV_vert}. With increasing treatment time, both the duration of the microdischarges and the peak current increase, which is consistent with previous research \cite{clyne_2018_review,Bracht_diss, hermanns_diss}. This effect is strongest for an Al substrate with 1\,g/l (see figure~\ref{res:Al_CV_vert} a)) and 2\,g/l of KOH (figure~\ref{sup:Al_U-I} b)), where the lifetime changes from below 50\,\textmu s to several hundred \textmu s, and the peak current from 20\,mA to 100\,mA. In contrast, this effect seems less strong for higher KOH concentration like in figure~\ref{res:Al_CV_vert}b). For example, the maximum current with an Al substrate drops from 93\,mA at 1\,g/l of KOH to 39\,mA at 3\,g/l of KOH (see figures~\ref{res:Al_CV_vert} and \ref{sup:Al_U-I}). Multiple measurements support the observation that the discharge current does not increase as much as it does in measurements with 1\,g/l of KOH.
Furthermore, the current profiles become less defined and show lower reproducibility at higher concentrations. Bracht \cite{Bracht_diss} similarly observed sharp and irregular current patterns at KOH concentrations around 4\,g/l. The irregular peaks become more prominent between 1\,g/l and 4\,g/l of KOH, with the most noticeable variation between 2\,g/l and 3\,g/l of KOH.
Generally, the increase in lifetime and maximum current peak per discharge pulse is caused by a growing oxide layer during the process, which results in more localized and less frequent microdischarges. Due to a constant current supply and since microdischarges are primarily responsible for the current through the oxide, an increase in the current per microdischarge leads to a reduction in ignition frequency. Furthermore, an increase in the probability of high current discharges can negatively affect the coating structure, potentially causing parts of the coating to detach. \\

Changing the substrate to Ti results in similar observations of increased microdischarge lifetime and higher current peaks over treatment time, as shown in figure~\ref{res:Ti_CV_vert} and in supplementary data figure~\ref{sup:Ti_U-I}. However, the peak current is significantly lower ($I_{max}\leq 25\,\mathrm{mA}$) compared to Al, making the effect difficult to observe. The effect of lower current and more frequent ignitions for an increased concentration of KOH is only visible with 0.5\,g/l of KOH, where typically one ignition occurs per ms and 1\,g/l of KOH with several ignitions. This effect is not observed at higher concentrations due to a high number of transient discharges as well as a very low peak current. A concentration of 2\,g/l of KOH shows more ignitions with lifetimes under 10\,\textmu s compared to 1\,g/l of KOH. However, significantly more data is required to determine whether the current is affected by such high concentrations.
Furthermore, compared to an Al substrate, a constant current of approximately 7 to 13\,mA is observed for all tested KOH concentrations, as seen in figures~\ref{res:Ti_CV_vert} and in supplementary data fiurge~\ref{sup:Ti_U-I}. This behaviour may be caused by a slower oxide layer growth, which is shown by the slow voltage increase over treatment time, as seen in figure~\ref{res:Ti_time02}. The difference in oxide layer conductivity between Ti and Al substrates could provide an additional explanation. It should be noted that the electrical conductivities ($\sigma$) of the oxide layers are difficult to compare without additional measurements due to dependency on temperature and the type of coating structure. As a first indication, the electrical conductivity of oxide coatings on Ti, such as TiO$_{2}$ (dielectric, $\sigma \approx 10^{-10}$\,S/m) and TiO (conductor, $\sigma \approx 10^{6}$\,S/m), is higher compared to Al$_{2}$O$_{3}$ (dielectric, $\sigma \approx 10^{-12}$\,S/m) \cite{CRC_handbook_material}.
The same explanation applies to the formation of the initial passivation layer, which also differs significantly between substrates. For Al, this layer forms within seconds, while on Ti it takes several seconds to minutes.
The slower oxide formation on Ti is reflected in the voltage behaviour seen in figures~\ref{res:Al_time02} and \ref{res:Ti_time02}, where the voltage is plotted over treatment time. While Al substrates show a voltage difference of 40\,V to 133\,V within 10 minute treatment time, Ti substrates exhibit a smaller difference of 10\,V to 33\,V. Interestingly, for Al the highest voltage difference observed during a single process occurs at the lowest concentration of 0.5\,g/l of KOH, whereas for Ti the highest difference is observed at 2\,g/l of KOH. The reason for this remains unclear, but it should be noted that the entire electrolyte/gas/oxide/substrate system can influence the voltage gap. \\

An increase in voltage over the treatment time, as most clearly observed with an Al substrate in figure~\ref{res:Al_time02}, indicates a thicker or denser coating structure over time. Since the voltage is measured across both the electrolyte and the coating, this interpretation is only valid if the electrolyte's conductivity remains nearly constant throughout the 10 minute treatment. Bracht \cite{Bracht_diss} confirmed this stability for an Al substrate, observing no significant changes in pH or conductivity over that period. Therefore, it is assumed that a rise in voltage indicates an increase in coating resistance. Bracht's study \cite{Bracht_diss} also highlighted differences in coating thickness between 1\,g/l (30\,-\,50\,\textmu m) and 4\,g/l (14\,-\,20\,\textmu m) of KOH for an Al substrate. It should be noted that the thickness in the present study are likely different from those reported by Bracht due to different current densities, which affect the oxidation rate of the process. A trend of decreasing thickness for higher KOH concentration would explain the lower voltage increase for 4\,g/l of KOH and an Al substrate. In Bracht's work \cite{Bracht_diss}, the coating thickness is measured by an SEM analysis. Further studies are needed to cover this aspect. \\

As mentioned before, concentrations of 2\,g/l of KOH and greater, with a Ti substrate, exhibit a completely different behaviour of current and voltage over treatment time. For instance, the voltage behaviour above 2\,g/l of KOH in figure~\ref{res:Ti_time02} shows relatively low voltages and large error bars compared to lower electrolyte concentrations. It was determined that the passivation layer does not become thick enough to build up a voltage before breakdown that is sufficient for microdischarge ignition. Instead, part of the coating detaches, causing a drop in resistance and voltage. This observation is additionally supported by the presence of coating residues in the liquid following a PEO treatment and an additional analysis with a SEM. \\

Typically, the voltage in the beginning of the PEO process decreases with higher KOH concentrations. For instance for an Al substrate from 535\,V at 0.5\,g/l to 453\,V at 4\,g/l of KOH (figure~\ref{res:Al_time02}). This voltage reflects the combined resistance of the oxide layer and the electrolyte solution. A higher electrolyte concentration increases the solution conductivity, thereby reducing the overall resistance. However, the reduction in voltage is not linear, suggesting additional factors. Similar trends are observed with Ti substrates, as shown in figure~\ref{res:Ti_time02}. These findings indicate that the concentration of the electrolyte also affects the resistance of the initial oxide interface. \\

By comparing the voltage and current behaviour of a PEO process for Al and Ti substrates, it was shown that Al substrates experience faster and thicker oxide layer growth than Ti. For Al, increasing the KOH concentration leads to thinner oxide layers, as shown by Bracht \cite{Bracht_diss}, and by the observed decreasing voltage change between the start and end of the process. Furthermore, an increase of KOH concentration leads to more indistinct and irregularly shaped current pulses. On Ti, the formation of an oxide coating is slower and no coating can be formed at KOH concentrations above 2\,g/l. Additionally, the number of ignitions increases with higher KOH concentrations, up to 2\,g/l, before ceasing entirely.

\subsection{Bubble dynamics}
\begin{figure}
    \centering
    \includegraphics[width=8cm]{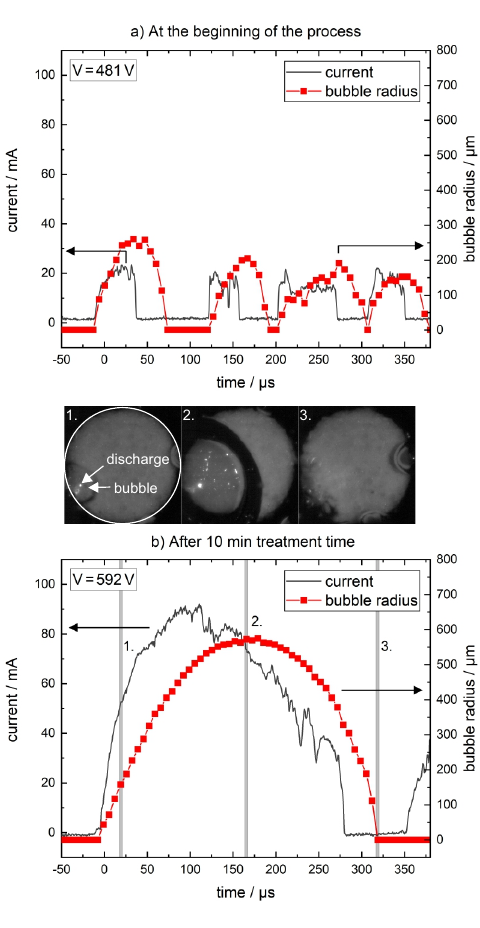}
    \caption{Time-resolved current measurement with the corresponding bubble radius for an aluminium substrate and 1\,g/l of KOH. Bubble formation is shown in the top images in b). The numbers correspond to the marked areas in the graphs below. a) shows the bubble at the beginning, and b) after 10 minutes of operation.} 
    \label{res:al_bubble_1gl_image}
\end{figure}

\begin{figure}[ht]
    \centering
    \includegraphics[width=8cm]{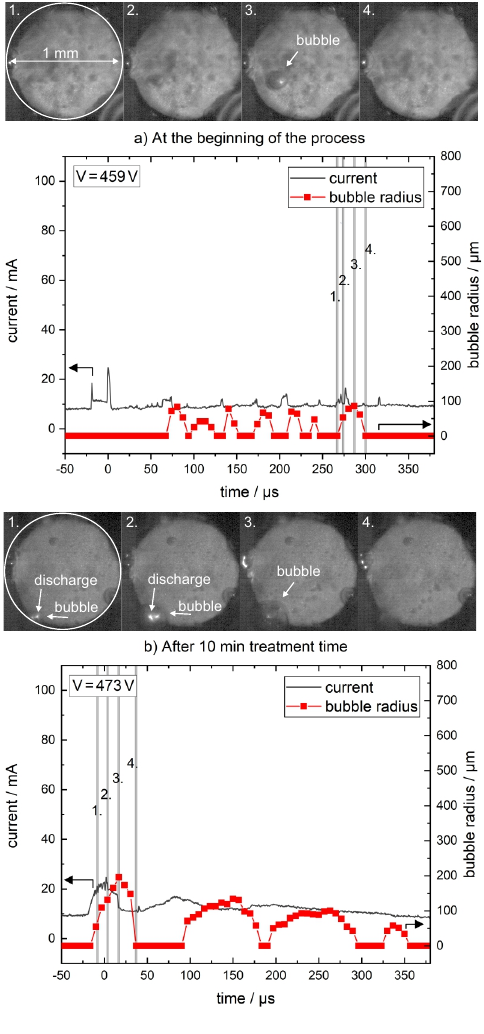}
    \caption{Time-resolved current measurement with the corresponding bubble images for a titanium substrate. Bubble growth and collapse are seen in the top images, where the numbers correspond to the highlighted time points in the graphs below. a) presents the beginning of the process for 1\,g/l of KOH and b) after 10 minutes of treatment time.} 
    \label{res:ti_bubble_1gl_image}
\end{figure}

As mentioned, the microdischarge lifetime increases over treatment time for both Al and Ti substrates. The bubble is generated at the same time as the microdischarge and begins to collapse as the microdischarge extinguishes. This indicates a direct relation between the microdischarge formation and bubble lifetime. In figure~\ref{res:al_bubble_1gl_image} and \ref{res:ti_bubble_1gl_image} discharge current (black) and bubble radius (red) are plotted for a duration of 380\,\textmu s starting from triggering on the microdischarge current at the start of the process a) and after 10 minutes of treatment time b). The electrolyte is composed of 1\,g/l KOH.
Over a 10 minute treatment of an Al substrate (figure~\ref{res:al_bubble_1gl_image}), bubble lifetimes increase from less than 100\,\textmu s to several hundreds of \textmu s, while maximum radii grow from 200\,–\,300\,\textmu m to over 550\,\textmu m. An exception is observed at 3\,g/l of KOH, where lower-current microdischarges result in a slight radius decrease (figure~\ref{sup:al_bubble_2gl-4gl_image}). Notably, at 4\,g/l of KOH no visible bubbles form after a 6-minute treatment time, likely due to a thick oxide layer preventing further microdischarges at the anode tip. Instead, voltage fluctuations and current flow with ignitions in non-visible areas are observed. That is consistent with the observations of the current behaviour for 3\,g/l and 4\,g/l of KOH (section~\ref{CV_section}). \\

Figure ~\ref{res:ti_bubble_1gl_image} shows the behaviour of current and bubble radius for Ti substrates at the start of the PEO process a), and after 10 minutes of treatment time b), again for 1\,g/l KOH. Here, fast bubble dynamics (10\,–\,35\,\textmu s initially, increasing to 50\,–100\, \textmu s) and small bubble radii ($\leq 100$\,\textmu m initially, slightly increasing to around 100\, \textmu m) are observed. Combined with a low image contrast, an accurate determination of bubble radius is challenging. Trends of increasing lifetime and peak radius can be determined, but are much lower compared to Al substrates. Furthermore, some bubbles could not be detected due to overlapping discharge events or noise in the current signal, as seen in figure~\ref{res:ti_bubble_1gl_image}. Determining the bubble radius was especially difficult at electrolyte concentrations above 2\,g/l KOH. \\

\begin{figure}
    \centering
    \includegraphics[width=8.5cm]{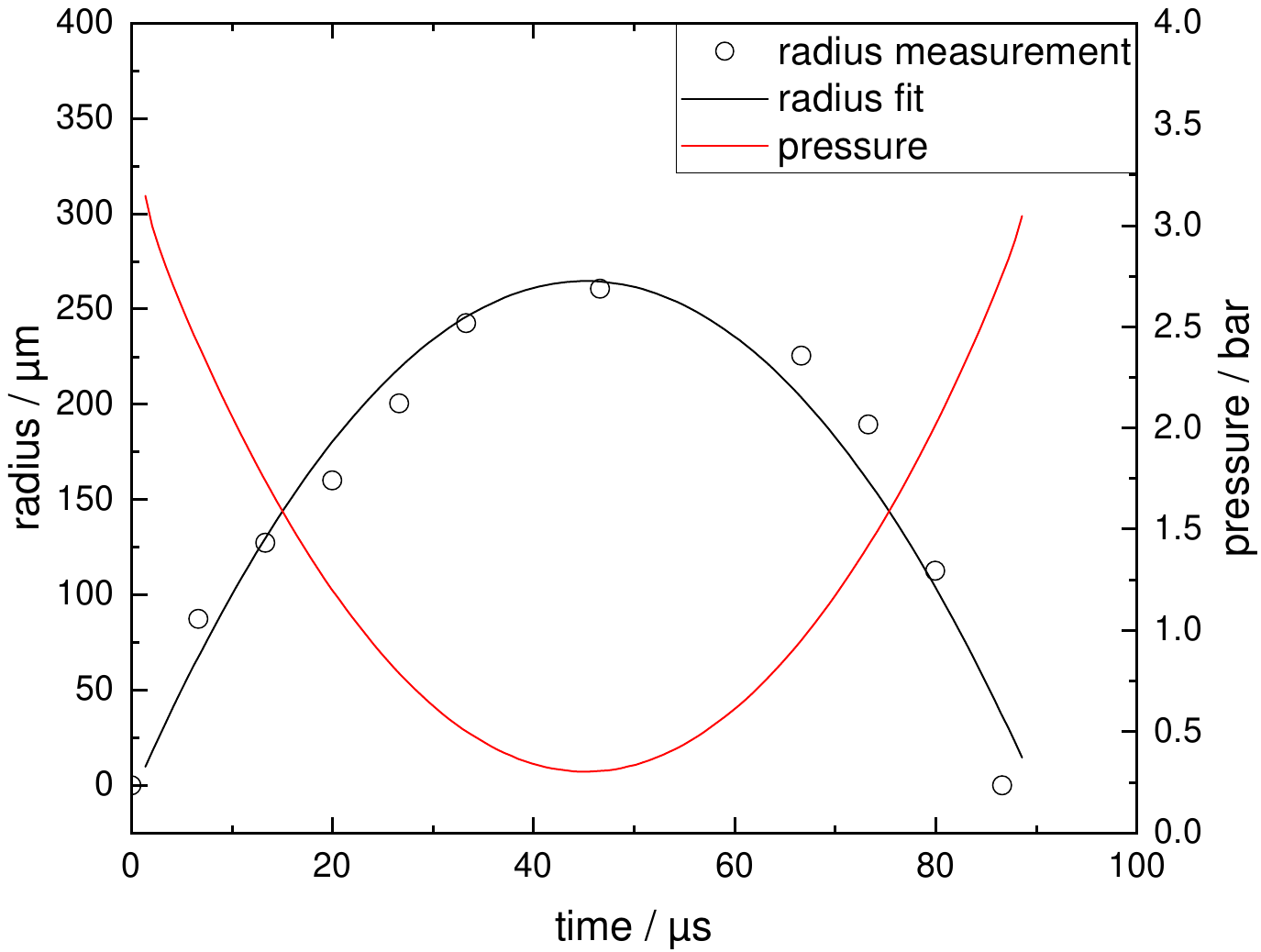}
    \caption{An example of the pressure profile for a bubble formed in 1\,g/l of KOH on an aluminium substrate. The bubble forms at the beginning of the process. The pressure is calculated using the Rayleigh-Plesset equation, with a second-degree polynomial fit for the radius as the input parameter.} 
    \label{Al_rayleigh}
\end{figure}

The bubble pressure is calculated using the Rayleigh-Plesset equation (see equation~\eqref{equ:rayleigh}), with the bubble radius as an input parameter. Figure~\ref{Al_rayleigh} illustrates the calculated pressure for an Al substrate in 1\,g/l KOH at the beginning of the PEO process, based on a second-degree polynomial fit of the bubble radius. The resulting pressure ranges between 0.5 and 3 bar, which is consistent with the findings of Bracht~\cite{Bracht_diss} and Troughton~\textit{et al.}~\cite{clyne_2015}. The pressure in the surrounding liquid is assumed to be atmospheric (p\,$\approx 1$\,bar). During the bubble growth phase, the pressure inside the bubble decreases from an overpressure state to an under pressure state, reaching its minimum at the maximum bubble radius. This low pressure within the bubble triggers a collapse, during which the pressure rises again above 1 bar.
Hamdan~\textit{et al.}~\cite{hamdanDynamicsBubblesCreated2013} proposed a much higher pressure at the start of the bubble growth, reaching 10s of bars in the first 100\,ns. Their study, however, used heptane instead of water and relied on a different non-PEO experimental setup, so only qualitative comparisons can be made with this work.
Due to the limitations of the temporal resolution of the camera used in this work (6.6\,\textmu s between each frame) such a phenomenon cannot be sufficiently resolved here. Further, the pressure behaviour appears to be relatively unaffected by the electrolyte concentration (figure~\ref{sup:Al_pressure}) or the substrate material (not shown), as it is primarily dependent on individual bubble dynamics. This finding aligns with the results reported by Bracht in \cite{Bracht_diss}, which also found no evidence of a pressure dependency on electrolyte concentration with an Al substrate. 
These observations highlight the correlation between bubble dynamics, microdischarges, and electrolyte properties during the PEO process. Both substrates typically show an increase in the average bubble lifetime and radius over treatment time. This means that a thicker oxide layer reduces the probability of a breakdown and therefore the microdischarge frequency decreases. However, this results in higher discharge intensity and, consequently, larger bubble formation.

\subsection{Electron and surface temperatures} 

\begin{figure}
    \centering
    \includegraphics[width=8.2cm]{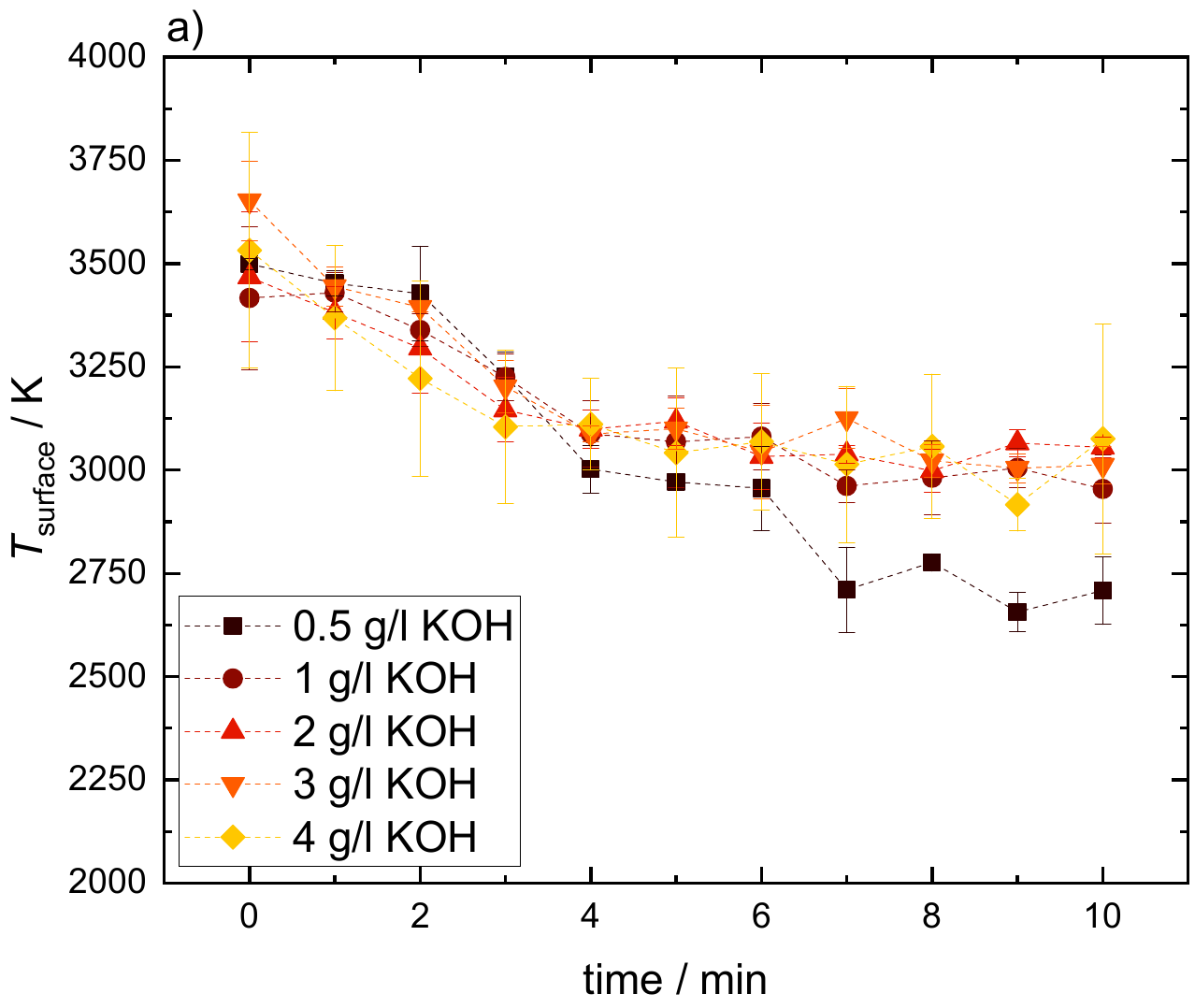}
    \includegraphics[width=8.2cm]{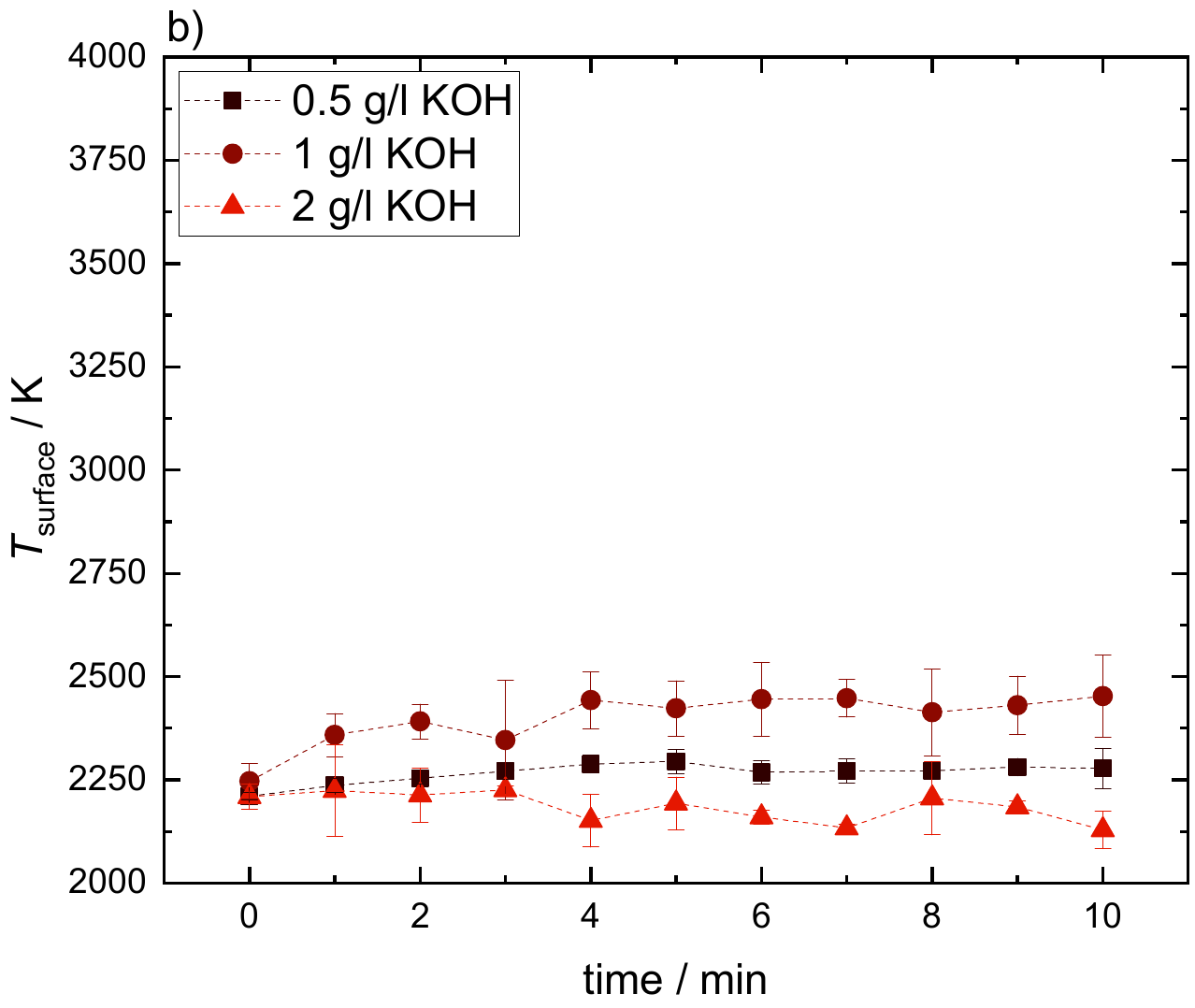}
    \caption{Calculated surface temperature and corresponding standard deviation for an aluminium substrate a) and a titanium substrate b) during the PEO process. It is obtained by fitting black body radiation to the measured continuum spectrum, based on several assumptions, such as the substrate behaving as a black body radiator.} 
    \label{Ts_Al_Ti}
\end{figure}

\begin{figure}
    \centering
    \includegraphics[width=8.2cm]{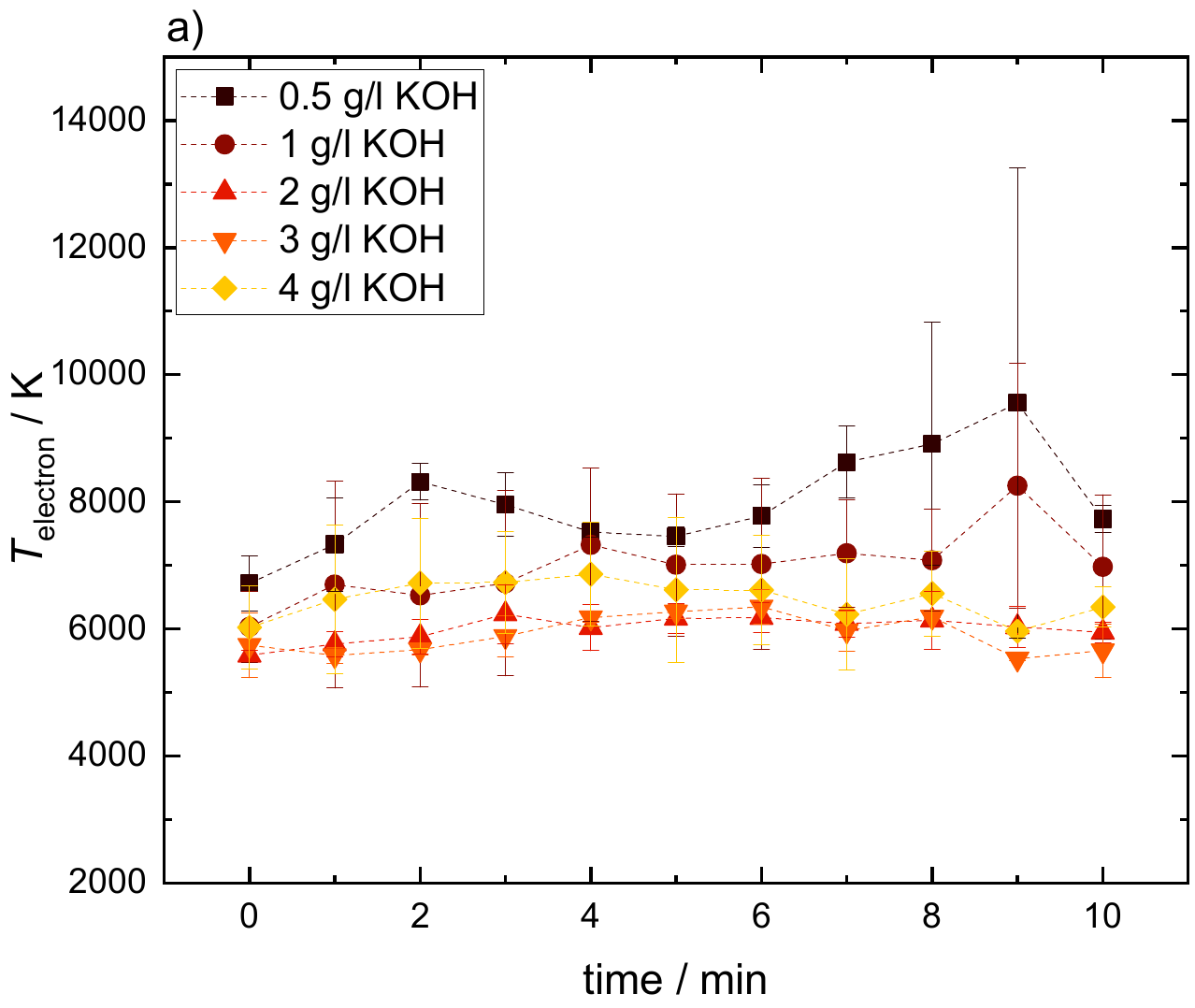}
    \includegraphics[width=8.2cm]{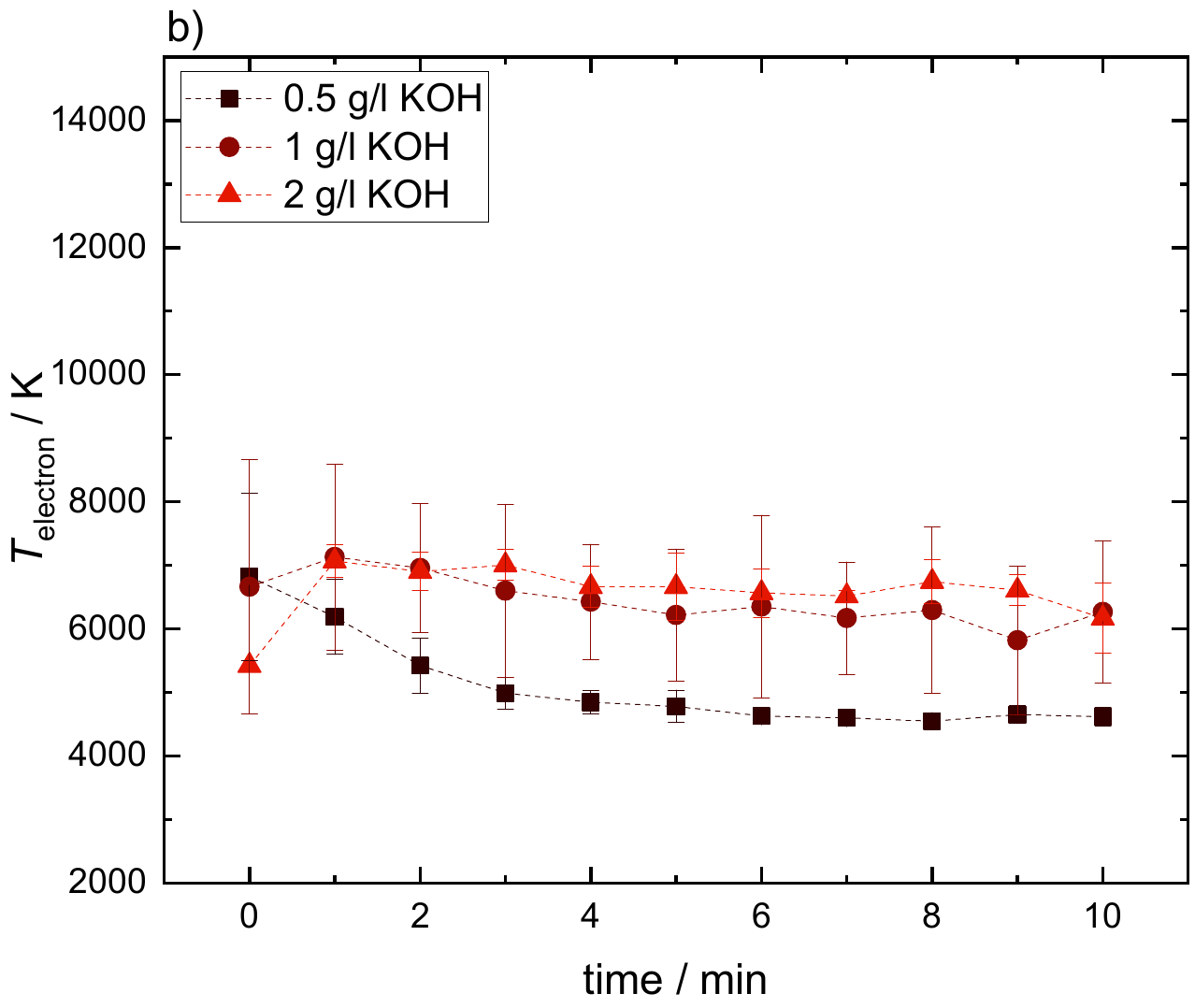}
    \caption{Calculated electron temperature and corresponding standard deviation for an aluminium substrate a) and a titanium substrate b) during the PEO process. It is obtained by fitting Bremsstrahlung to the measured continuum spectrum.} 
    \label{Te_Al_Ti}
\end{figure}

The surface and electron temperatures are calculated by fitting expressions for Bremsstrahlung (electron temperature) and black body radiation (surface temperature) to the measured continuum spectrum, as seen in the supplementary data figure~\ref{sup:cont_Al-Ti}. The estimates for the surface temperature over a treatment time of ten minutes are presented in figure~\ref{Ts_Al_Ti} with an Al a) and Ti substrate b), respectively.
The surface temperatures estimated with this method range from 2600\,K to 3750\,K on the Al anode and from 2000\,K to 2500\,K on the Ti anode. The estimated surface temperatures are mostly consistent with the observed thermal behaviour of the materials. As the layer undergoes localised melting and solidification during the process, it is expected to exceed the melting temperature of Al\textsubscript{2}O\textsubscript{3} at 2345\,K and 2116\,K for TiO\textsubscript{2}. This behaviour aligns with the findings of Hermanns~\textit{et al.}~\cite{hermanns_-situ_2020} and Bracht \cite{Bracht_diss}, who observed a similar phenomenon with an Al substrate.
The temperature on the Al substrate, as illustrated in figure~\ref{Ts_Al_Ti} a), shows a trend of decreasing surface temperatures over treatment time for all tested KOH concentrations, with a maximum of $\Delta T_s$\,=\,840\,K for 0.5\,g/l over 10 minutes of treatment. In contrast, this trend is not visible for a Ti substrate, moreover in the first minutes the temperature (0.5\,g/l KOH, 1\,g/l KOH) seems to be increasing with a much smaller difference of $\Delta T_s$\,=\,200\,K and reaches a more constant value after few minutes, as shown in figure~\ref{Ts_Al_Ti} b).
Based on current and voltage measurements for Al and Ti, an increase in surface temperature might be expected due to the generation of more local, high-current discharges over time. However, the findings indicate that the surface temperature does not rise over the treatment and rather decreases for an Al substrate.
The distribution of hot and cold spots on the substrate surface might explain these trends. An increase in high-current discharges lowers the discharge frequency, resulting in fewer but more intense discharges. However, these intense discharges seem to transfer heat less efficiently to the substrate compared to more frequent, lower-intensity discharges.
The estimation of electron temperatures using Bremsstrahlung fitting shows a high standard deviation, reaching 3700\,K, as seen in figure~\ref{Te_Al_Ti}. This is due to challenges in the fitting approach around 300\,nm, where the spectrum has a low intensity while the Bremsstrahlung has the most impact. The low intensity of Bremsstrahlung makes it difficult to determine whether to adjust the electron temperature or the scaling factor, leading to high errors for electron temperatures. Furthermore, the inaccuracy of calculating Bremsstrahlung could be due to the simplification in the fitting procedure. As an example, a different EEDF shape than a Maxwellian could increase the influence of Bremsstrahlung at around 500\,nm to 900\,nm, which would reduce the impact of black body radiation in that region. Additionally, black body radiation could have a greater effect at wavelengths above 900 nm. However, due to the spectrometer's sensitivity limitations in this range, its impact could not be accurately determined. \\

The electron temperature for an Al substrate is estimated to be between 5000\,K\,-\,12200\,K, as seen in figure~\ref{Te_Al_Ti} a). This temperature is notably higher than the surface temperature and aligns with findings from previous studies \cite{Bracht_diss, husseinEffectCurrentMode2019,jovovicSpectroscopicStudyPlasma2012,dunleavyCharacterisationDischargeEvents2009}. However, the error bars are large ($\le 3700$\,K) compared to the surface temperature. No clear influence of the temperature over treatment time is visible, though a slight trend to decreasing electron temperature with higher KOH concentration is noticeable.
The electron temperature for a Ti substrate is in a range of 4300\,K to 8600\,K, as seen in figure~\ref{Te_Al_Ti}b), and is similarly influenced by larger errors compared to the surface temperature, but less than for the Al anode. Similar to an Al substrate, an influence of the electrolyte concentration or the treatment time on the temperature is hard to identify due to large error bars (($\le 2000\,K$)), especially for a concentration of 2\,g/l of KOH. At 0.5\,g/l KOH, the temperature shows a decreasing trend over treatment time from 7000\,K to 4500\,K. The temperatures are expected to be higher than the surface temperature and are in a reasonable range, similar to an Al anode. \\

Overall, it could be shown that discharges cause a lower ($\Delta T\leq$1300\,K) surface temperature with a Ti substrate compared to Al. The temperatures are mostly in a reasonable range, but trends are difficult to identify. In contrast, a wide range of electron temperatures are calculated (4300\,–\,12200\,K), which are accompanied by high error bars showing the limits of the method used. However, since each measurement averages over hundreds of discharges and different discharges occur as proposed by Hussein et al.~\cite{husseinInvestigationCeramicCoating2013}, the observed temperature variations are reasonable.

\subsection{Morphology of the coating}
\begin{figure}
    \centering
    \includegraphics[width=8.2cm]{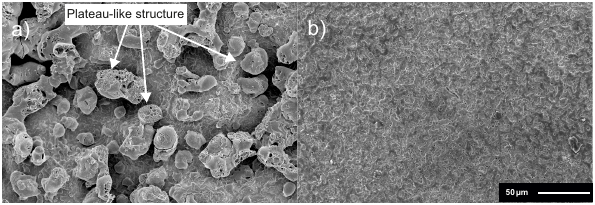}
    \caption{SEM images at 450x magnification showing the inner layer of Al (left) and Ti (right) after a 10 minute PEO treatment with 1\,g/l\,KOH. For Ti, no combination of inner and outer layers was observed.}
    \label{SEM:Al-Ti}
\end{figure}

\begin{figure}
    \centering
    \includegraphics[width=8.2cm]{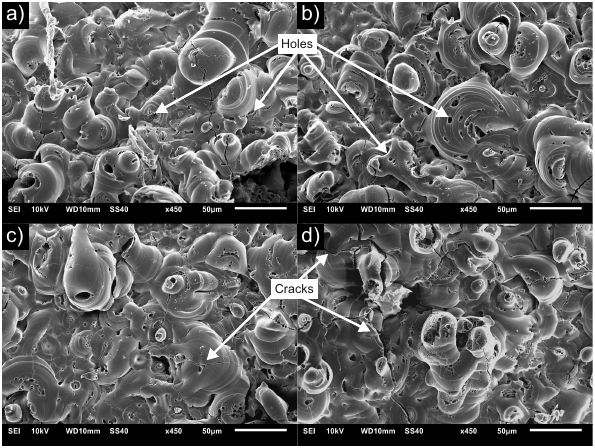}
    \caption{SEM images with a magnification of 450x for Al substrates and different KOH concentrations: a) 1\,g/l\,KOH, b) 2\,g/l\,KOH, c) 3\,g/l\,KOH and d) 4\,g/l\,KOH. The images show the outer layer of the coating.} 
    \label{SEM:Al}
\end{figure}

The coating morphology after a 10 minute PEO treatment differs significantly between an Al and a Ti substrate, as shown in figure~\ref{SEM:Al-Ti}. On Al, the PEO process generates a multi-layer structure consisting of an inner and outer layer, which is in agreement with Hussein~\textit{et al.}~\cite{husseinInvestigationCeramicCoating2013}. The inner layer is a thin, compact layer formed at the interface between the substrate and the outer layer. The thickness of the outer layer is typically around 10\,\textmu m. The maximum thickness of the inner layer was determined to be in the range of 5\,-\,10\,\textmu m by Bracht \cite{Bracht_diss} using half of the current density at the anode under the same conditions. Furthermore, Bracht showed a minimum thickness of the inner contact layer to be below 1\,\textmu m. This aspect is not further investigated in this study. However, the outer layer is prone to damage during the sample preparation, such as removing the fluorine rubber O-ring after the PEO process. This often leads to partial detachment of the outer coating layer, which is mostly observed for a concentration of 1\,g/l of KOH. Figure~\ref{sup:SEM:Al_layers} shows this effect, where both layers can be seen. This indicates weaker adhesion compared to coatings formed at higher KOH concentrations, where the outer layer remains more intact.
For an Al substrate, the outer layer, as seen in figure~\ref{SEM:Al}, is smoother, with less holes and graininess, while the inner coating layer contains many plateau-like regions with diameters in the range of 20\,\textmu m, as shown in figure~\ref{SEM:Al-Ti} a). These structures are connections from the inner to outer layer. The holes are typically in the range of 0.5\,-\,10\,\textmu m, though both larger and smaller holes can also be observed. In addition, the outer coating exhibits large cracks, often extending tens to hundreds of micrometers in length with a typical width of approximately 1\,\textmu m. Overall, the morphology at different KOH concentrations is very similar, and more data is needed to show if smaller differences can be distinguished. However, as mentioned the analysis by Bracht \cite{Bracht_diss} showed that the coating thickness is decreasing for higher electrolyte concentration. \\

\begin{figure}
    \centering
    \includegraphics[width=8.2cm]{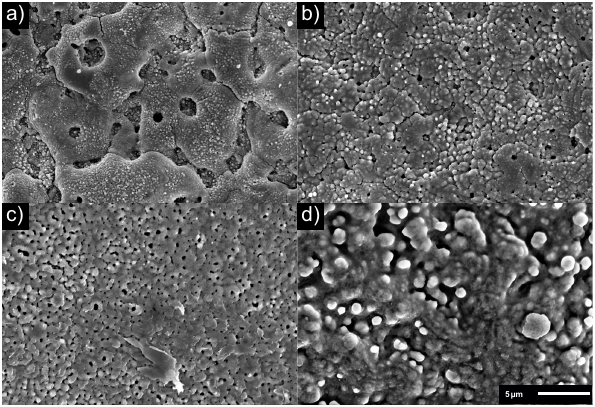}
    \caption{SEM images with a magnification of 4500x for Ti substrates and different KOH concentrations: a) 1\,g/l\,KOH, b) 2\,g/l\,KOH, c) 3\,g/l\,KOH and d) 4\,g/l\,KOH.} 
    \label{SEM:Ti}
\end{figure}

In contrast, Ti coatings exhibit a large number of much smaller pores ($\leq$\,2\,\textmu m), as seen in figure~\ref{SEM:Ti} a), forming a denser passivation layer compared to Al. Unlike Al, no signs of degradation of an outer layer is observed, making it impossible to distinguish between an outer and an inner layer. However, further studies are necessary to confirm this observation. Furthermore, the variation of the KOH concentration shows that above 2\,g/l of KOH no PEO coating is generated, as shown in figure~\ref{SEM:Ti}. Moreover, it seems that the surface layer is more likely modified by an anodising process and less thick compared to coatings formed with a KOH concentration of 1\,g/l. This is in agreement with the voltage and current measurements, which showed significant fluctuations for more than 2\,g/l of KOH and an absence of microdischarge ignitions. \\

EDX analysis, as presented in table~\ref{tab:EDX_AL_Ti}, confirmed the formation of TiO$_{2}$ on Ti and Al$_{2}$O$_{3}$ on Al when treated with KOH electrolyte. In addition, some amount of gold was measured due to the gold coating during preparation and sometimes small amounts of potassium were found. The EDX results are consistent for all tested KOH concentrations (1\,-\,4\,g/l KOH), even if no PEO ignition could be observed (Ti: 3\,-\,4\,g/l KOH). However, Ti surfaces treated with 2 and 3\,g/l of KOH exhibit brightness variations in the SEM images, which most likely indicates differences in passivation layer density. EDX analysis on these spots further revealed a uniform TiO$_{2}$ content, indicating that darker areas are denser and less porous than brighter regions. \\

These findings show the importance of tailoring the PEO parameters to achieve desired coating properties. Under otherwise similar KOH electrolyte concentrations, Ti substrates generally form more compact and uniform layers, with presumably lower coating thickness in comparison to Al substrates. PEO on both substrates is affected by the electrolyte concentration. However, due to the small sample size for Al and variations of the inner and outer layer, which is sometimes not visible, no significant effect of KOH concentration on the morphology could be observed. In contrast, coating morphology on Ti is highly affected by electrolyte concentration, which is also reflected by the electrical measurements. For Ti substrates, a typical PEO structure characterized by a large number of small pores is formed at 1\,g/l of KOH, while higher concentrations result in variations in coating density and less PEO-specific characteristics and a transition to what appears to be an anodising process.

\begin{table}[]
\centering
\begin{tabular}{@{}cccc@{}}
\toprule
\multicolumn{2}{c}{Titanium substrate} & \multicolumn{2}{c}{Aluminium substrate} \\ \cmidrule(lr){1-2} \cmidrule(lr){3-4}
\makebox[3em]{\shortstack{\vspace{5pt} Element}} & \makebox[6em]{\shortstack{Atom \\ (at. \%)}} & \makebox[3em]{\shortstack{\vspace{5pt} Element}} & \makebox[6em]{\shortstack{Atom \\ (at. \%)}} \\ \midrule
O     & 64.83   & O   & 57.89 \\
Ti    & 32.32   & Al  & 38.59 \\
Au    & 2.47    & Au  & 3.52  \\
K     & 0.39    & -   & - \\
\bottomrule
\end{tabular}
\caption{Composition of the coating layer after a 10 minute PEO treatment of Ti and Al substrates. The results are obtained by EDX. To ensure conductivity the tip is coated with Au. The amount of the substrate and oxygen indicates the formation of TiO$_{2}$ or Al$_{2}$O$_{3}$.}
\label{tab:EDX_AL_Ti}
\end{table}

\section{Conclusion and outlook}
The present study has investigated the influence of different KOH electrolyte concentrations on microdischarge behaviour during plasma electrolytic oxidation (PEO) of aluminium (Al) and titanium (Ti) substrates. By employing a single microdischarge setup along with diagnostic tools such as high-speed imaging, optical spectroscopy and scanning electron microscopy, the impact of microdischarge properties on the PEO process and the coating could be analysed.
The results revealed significant differences in the PEO processes for these substrates, for otherwise similar conditions, as well as the KOH electrolyte concentration. Al substrates showed faster oxide layer growth and thinner coatings with higher KOH concentrations, indicated by a reduced voltage gap during process and supported by coating measurements by Bracht \cite{Bracht_diss}. In contrast, coating formation on Ti substrates was slower, with no PEO coating forming at KOH concentrations above 2\,g/l. Increasing the KOH concentrations result in an increased number of microdischarge ignitions on Ti up to 2\,g/l, while on Al it results in more indistinct and irregular shaped discharges. Furthermore, a constant current flow of several mA and much lower current peaks are observed during the PEO process of Ti compared to Al. \\

For both substrates, bubble lifetimes and radii increased over treatment time, which are correlated with microdischarge intensity and their lifetimes, due to a growing oxide layer. Furthermore, the growing oxide layer results in reduced microdischarge frequency, causing intensified individual discharges.
The calculated surface temperatures for Ti were approximately 1000\,K lower compared to Al, and were in the range of 2000\,K to 2500\,K, though trends were challenging to discern. These temperatures are typically above the melting temperature of TiO$_2$ (2116\,K), confirming the observed melting and solidification of the oxide. Electron temperature estimates ranged widely (4000\,–\,12200\,K), with significant error margins, highlighting an oversimplification of the approach with the used method. \\

Finally, morphological analysis showed the importance of optimizing PEO parameters to achieve desired coating properties. Ti substrates formed compact and uniform oxide layers with reduced thickness compared to Al substrates. The Al coating consists of a multilayered structure, with the morphological differences being minimally influenced by the KOH electrolyte concentration. However, additional SEM images are necessary to reveal morphology differences, and according to the study by Bracht \cite{Bracht_diss}, the coating thicknesses vary significantly. Coatings on Ti substrates exhibited a strong dependency of the electrolyte concentration on the morphology, with a typical PEO pore structure forming at 1\,g/l of KOH, while higher concentrations resulted in no PEO specific coatings.

\section*{Acknowledgement}
This study was funded by the German Research Foundation (DFG) with the Collaborative Research Centre CRC1316 ``Transient atmospheric plasmas: from plasmas to liquids to solids" (project B5).

\section{ORCID iDs}
Jan-Luca Gembus: https://orcid.org/0009-0002-1263-9218 \\
Vera Bracht: https://orcid.org/0000-0003-4623-7532 \\
Lars Schücke: https://orcid.org/0000-0002-7991-853X \\
Peter Awakowicz: https://orcid.org/0000-0002-8630-9900\\
Andrew R. Gibson: https://orcid.org/0000-0002-1082-4359

\section{Referencing\label{except}}
	
\bibliographystyle{ieeetr}  
\addcontentsline{toc}{chapter}{Bibliography}
\bibliography{bibliography}  

\pagebreak
\newpage
\onecolumn
\begin{center}
    \textbf{\large Supplementary information: Characterisation of single microdischarges during PEO of Al and Ti}
\end{center}
\setcounter{equation}{0}
\setcounter{figure}{0}
\setcounter{table}{0}
\setcounter{page}{1}
\setcounter{section}{0}
\makeatletter
\renewcommand{\theequation}{S\arabic{equation}}
\renewcommand{\thefigure}{S\arabic{figure}}
\renewcommand{\thetable}{S\arabic{table}}
\renewcommand{\thesection}{S\arabic{section}}
\renewcommand{\thepage}{S\arabic{page}}

\renewcommand{\theHequation}{S\arabic{equation}}
\renewcommand{\theHfigure}{S\arabic{figure}}
\renewcommand{\theHtable}{S\arabic{table}}
\renewcommand{\theHsection}{S\arabic{section}}
\makeatother
This supplementary data provides additional information supporting the findings presented in the main manuscript on plasma electrolytic oxidation (PEO). Aluminium (Al) and titanium (Ti) substrates are used in a single microdischarge setup to investigate the effect on individual microdischarges. It contains more detailed optical and SEM measurements with more data for different electrolyte concentrations of KOH (0.5 - 4\,g/l). \\

\section{Bubble and microdischarge formation}
\begin{figure*}[ht]
    \centering
    \includegraphics[width=5cm]{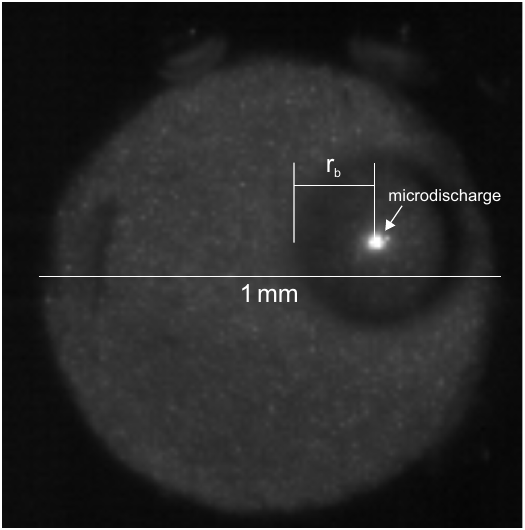}
    \caption{Top-view image of a bubble generated at the tip of a 1\,mm diameter Al wire. The bright spot represents a microdischarge surrounded by a bubble. The bubble radius, $r_{b}$, is measured in each frame captured during its formation.} 
    \label{sup:bubble}
\end{figure*}

Figure~\ref{sup:bubble} presents a top-view image of a bubble formed at the tip of a 1\,mm diameter Al wire. The bright spot within the image corresponds to a microdischarge, which is surrounded by a gas bubble. The bubble radius $r_{\mathrm{b}}$ is determined from each frame captured during its formation process. \\
\begin{figure*}[]
    \centering
    \includegraphics[width=17cm]{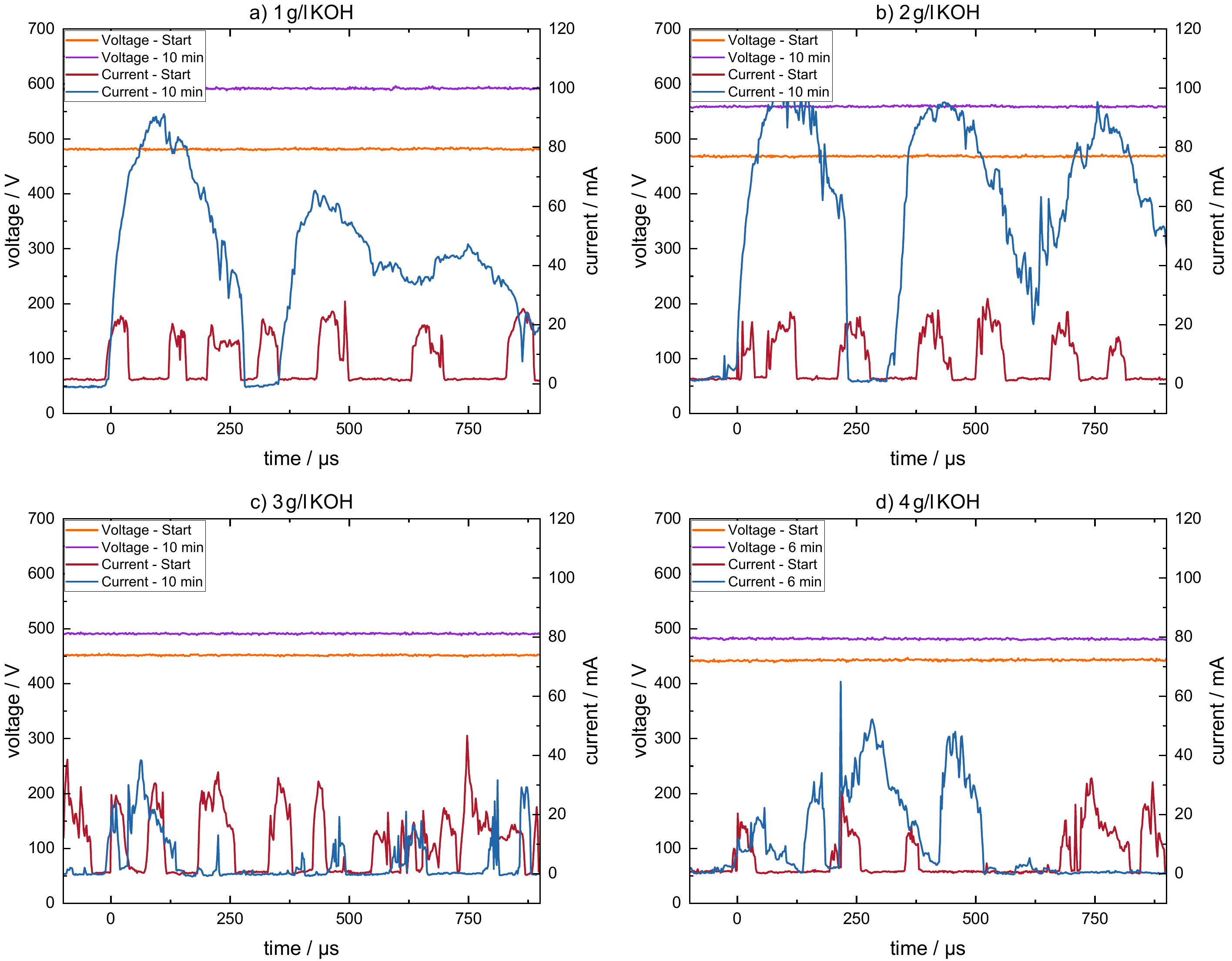}
    \caption{Time-resolved current-voltage measurements during the initial and final stages of a PEO treatment with an aluminium substrate. The voltage is shown in orange (start), purple (10\,min) and the current in red (start), blue (10\,min). Graphs a) to d) show the effect of increasing electrolyte (KOH) concentration. The treatment duration is 10 minutes, except in graph d), where it was 6 minutes.} 
    \label{sup:Al_U-I}
\end{figure*}

\begin{figure*}[]
    \centering
    \includegraphics[width=17cm]{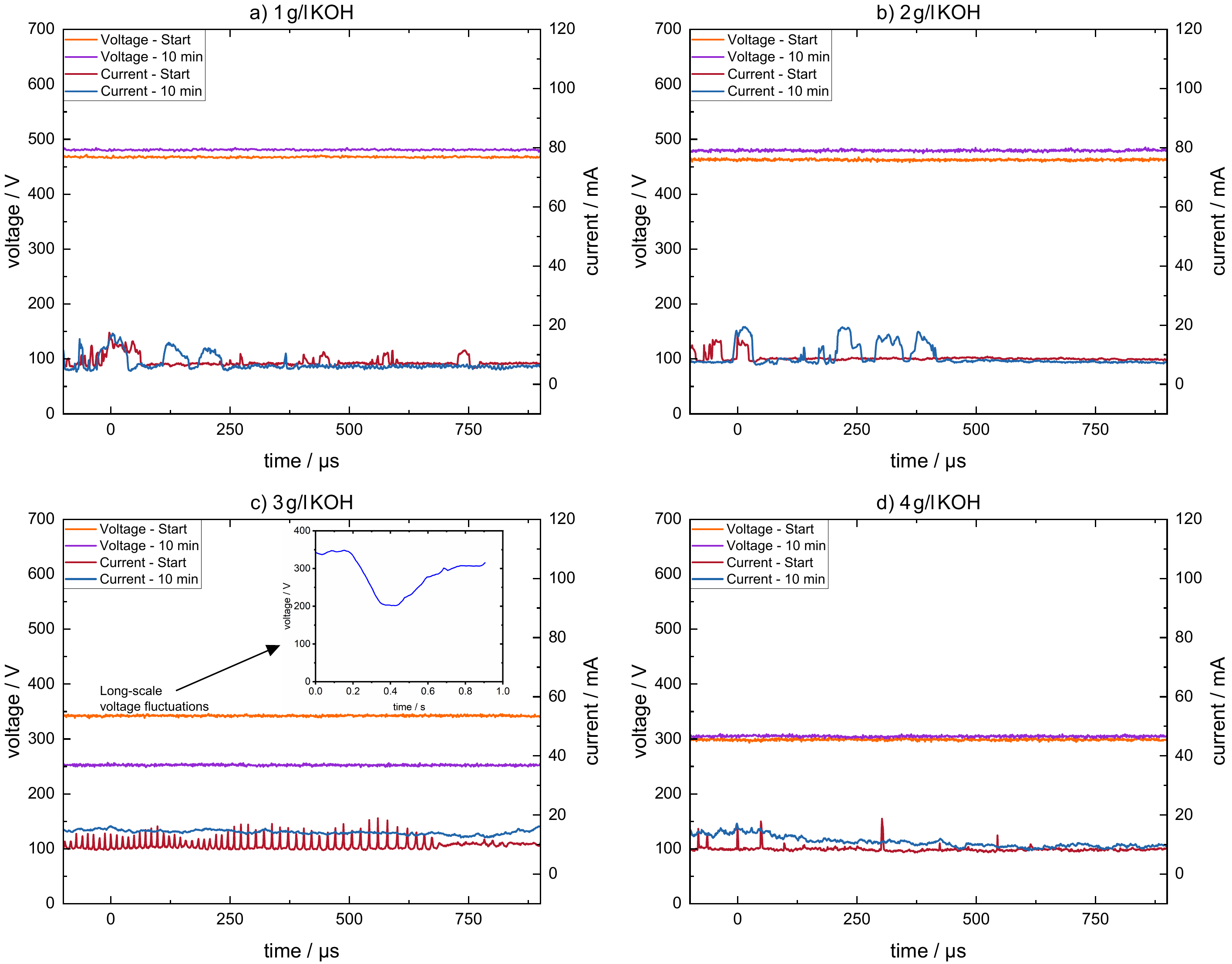}
    \caption{Time-resolved current-voltage measurements at the start and after 10 minute PEO treatment with a titanium anode. The voltage is shown in orange (start), purple (10\,min) and the current in red (start), blue (10.\,min). a) to d) shows the variation of the increasing electrolyte concentration of KOH. The voltage behaviour in c) and d) is not reproducible due to voltage fluctuations as seen in the graph c).} 
    \label{sup:Ti_U-I}
\end{figure*}

Figures~\ref{sup:Al_U-I} and \ref{sup:Ti_U-I} show time-resolved current-voltage measurements at the beginning and end of a PEO treatment with an Al and Ti anode. The treatment duration is 10 minutes, except for 4\,g/l of KOH with Al, where it was 6 minutes, after which the process stops by itself. Voltage is shown in orange (start) and purple (10\,min), while current is in red (start) and blue (10\,min). Graphs a) to d) illustrate the effect of increasing KOH concentration on the electrical parameter. \\

\begin{figure}[ht]
    \centering
    \includegraphics[width=8.2cm]{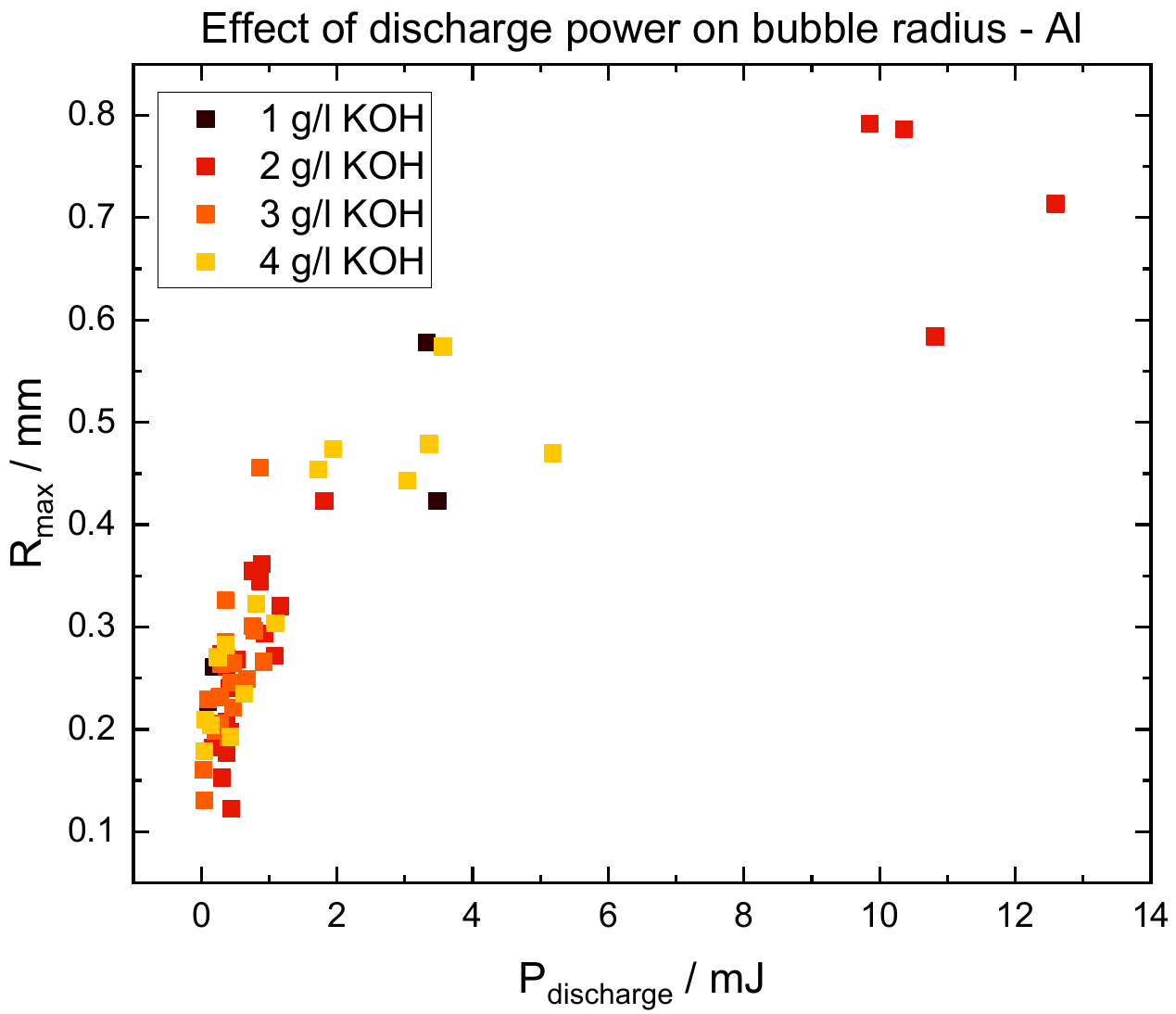}
    \caption{Scatter plot for the PEO process with Al substrate and different KOH concentration. Power is calculated by integrating the current of a microdischarge times the voltage.} 
    \label{sup:Al_scatter_power}
\end{figure}

Scatter plots for the PEO process with Al substrate and different KOH concentrations are shown in figure~\ref{sup:Al_scatter_power}. Exploring a power correlation with bubble dynamics could provide valuable insights. Generally, a larger bubble radius tends to correlate with greater power dissipation per microdischarge. However, distinguishing trends for different KOH concentrations is challenging due to the presence of outliers that do not follow the expected correlation. Additional data may be required to identify a clear trend. \\

\clearpage
\section{Investigation of bubble pressure and radius}
\begin{figure*}[ht]
    \centering
    \includegraphics[width=17cm]{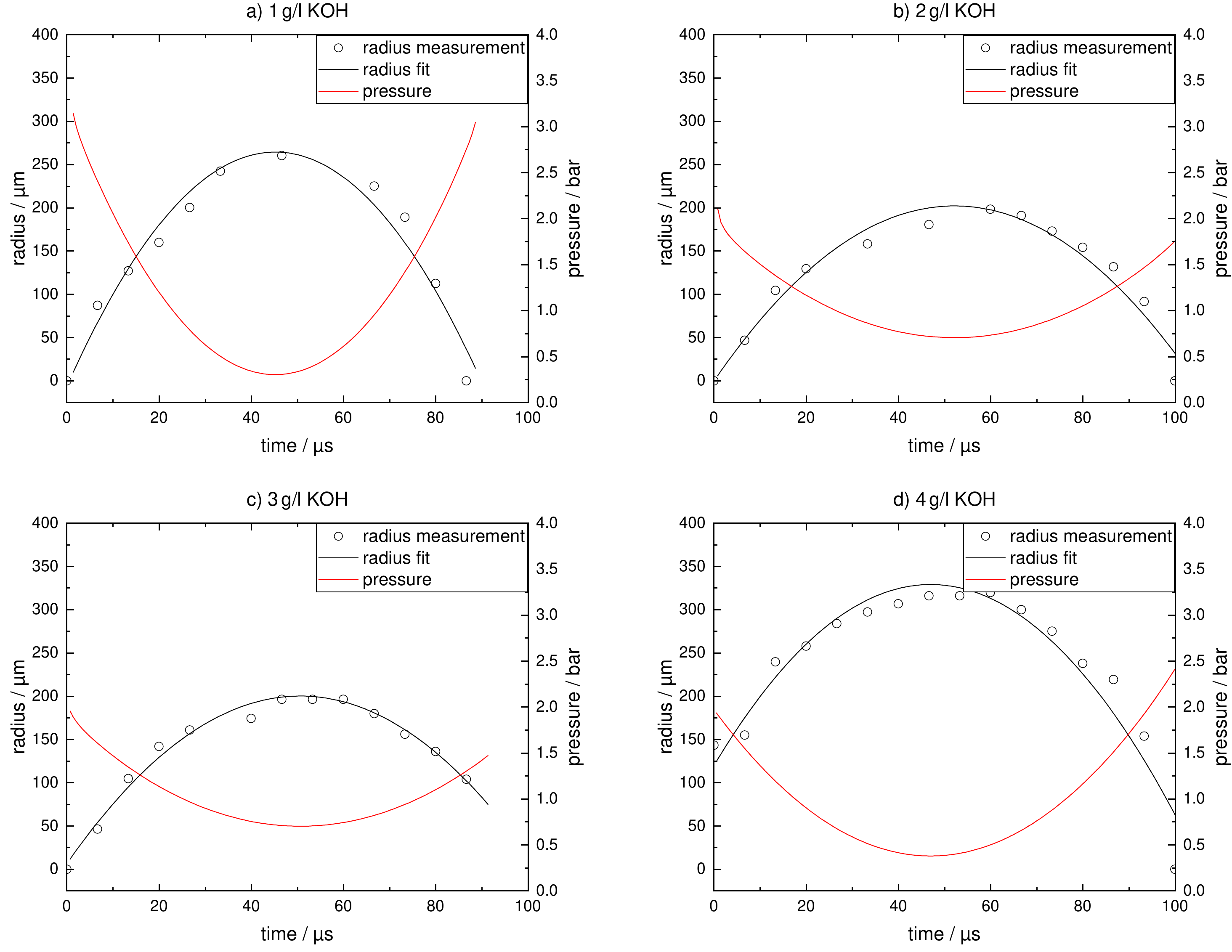}
    \caption{Examples of pressure profiles for bubbles formed in different electrolyte concentrations on an aluminium substrate. The pressure is calculated using the Rayleigh-Plesset equation, with a second-degree polynomial fit for the radius as the input parameter.} 
    \label{sup:Al_pressure}
\end{figure*}

Figure~\ref{sup:Al_pressure} illustrates examples of pressure profiles for bubbles generated in different electrolyte concentrations on an Al substrate. The pressure is derived using the Rayleigh-Plesset equation, employing a second-degree polynomial fit of the bubble radius as the input parameter. \\

\begin{figure*}[ht]
    \centering
    \includegraphics[width=17cm]{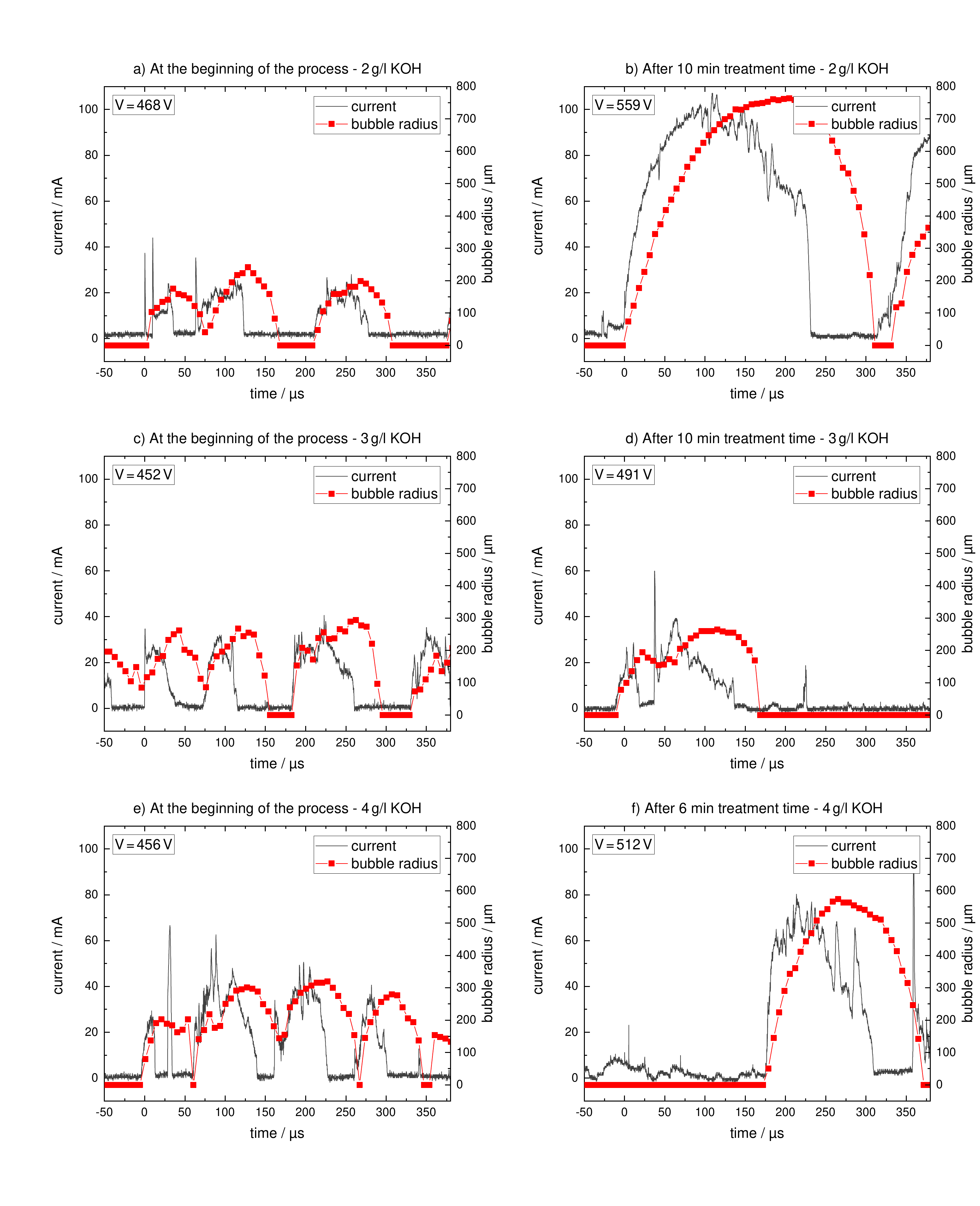}
    \caption{Time-resolved current measurement with the corresponding bubble dynamics for an aluminium anode and different amounts of KOH. a), c), e) show the dynamics at the beginning and b), d) after 10 minutes of the process. For 4\,g/l the process finished after 6 minutes treatment time, which is is shown in f).} 
    \label{sup:al_bubble_2gl-4gl_image}
\end{figure*}

Figure~\ref{sup:al_bubble_2gl-4gl_image} presents time-resolved current measurements with the corresponding bubble dynamics for an Al anode at varying KOH concentrations. Graphs a), c), and e) present the dynamics at the start of the process, while graphs b) and d) show the dynamics after 10 minutes. For a KOH concentration of 4 g/l, the process concluded after 6 minutes, as illustrated in graph f). \\

\begin{figure*}[ht]
    \centering
    \includegraphics[width=17cm]{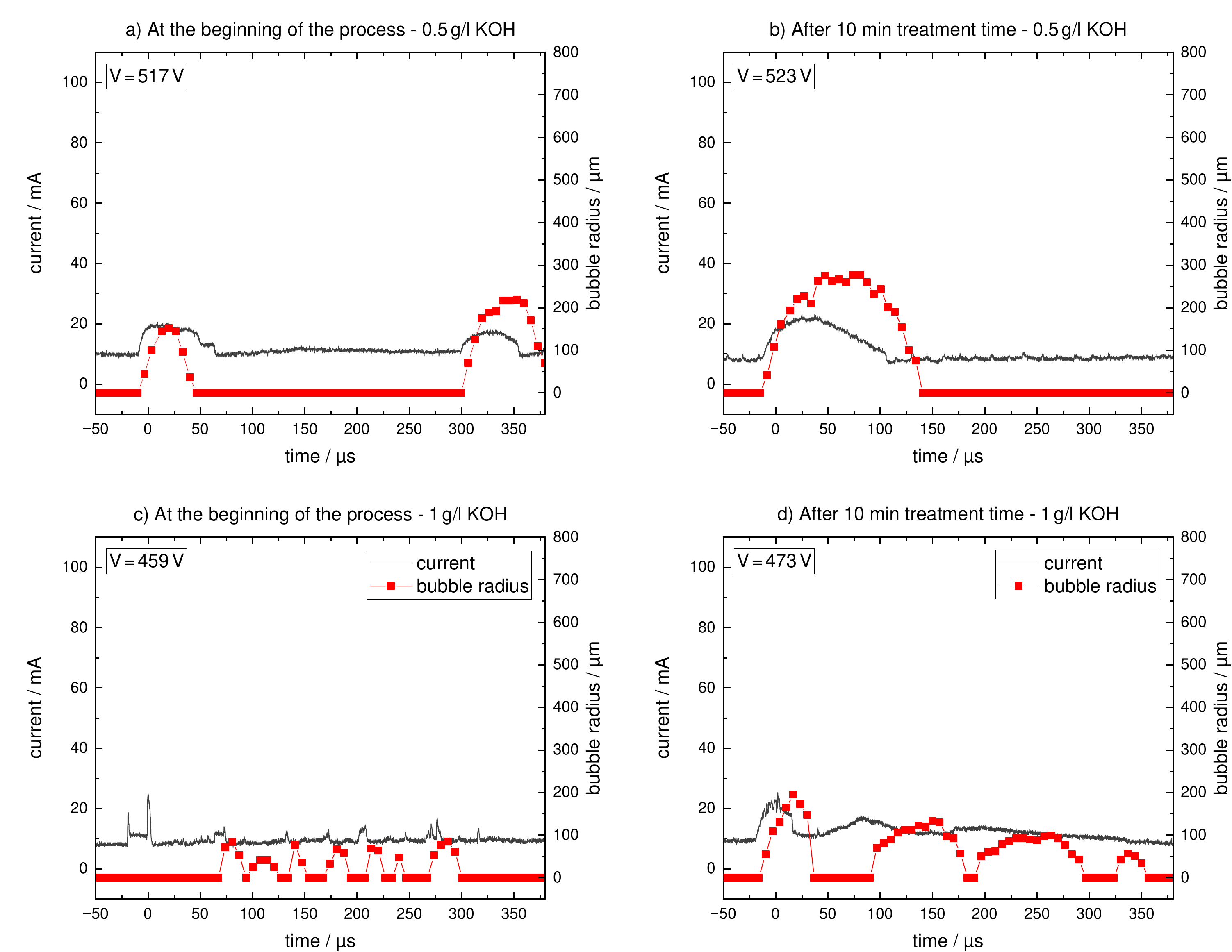}
    \caption{Time-resolved current measurement with the corresponding bubble dynamics for a titanium anode and different amounts of KOH. a), c) show the dynamics at the beginning and b), d) after 10 minutes of the process.} 
    \label{sup:ti_bubble_05gl-1gl_image}
\end{figure*}

Time-resolved current are shown~\ref{sup:ti_bubble_05gl-1gl_image} measurements with the corresponding bubble dynamics for a titanium anode at varying KOH concentrations are shown. Graphs a) and c) illustrate the dynamics at the start of the process, while graphs b) and d) represent the dynamics after 10 minutes. \\
\clearpage

\section{Spectroscopic analysis}
\begin{figure*}[ht]
    \centering
    \includegraphics[width=8cm]{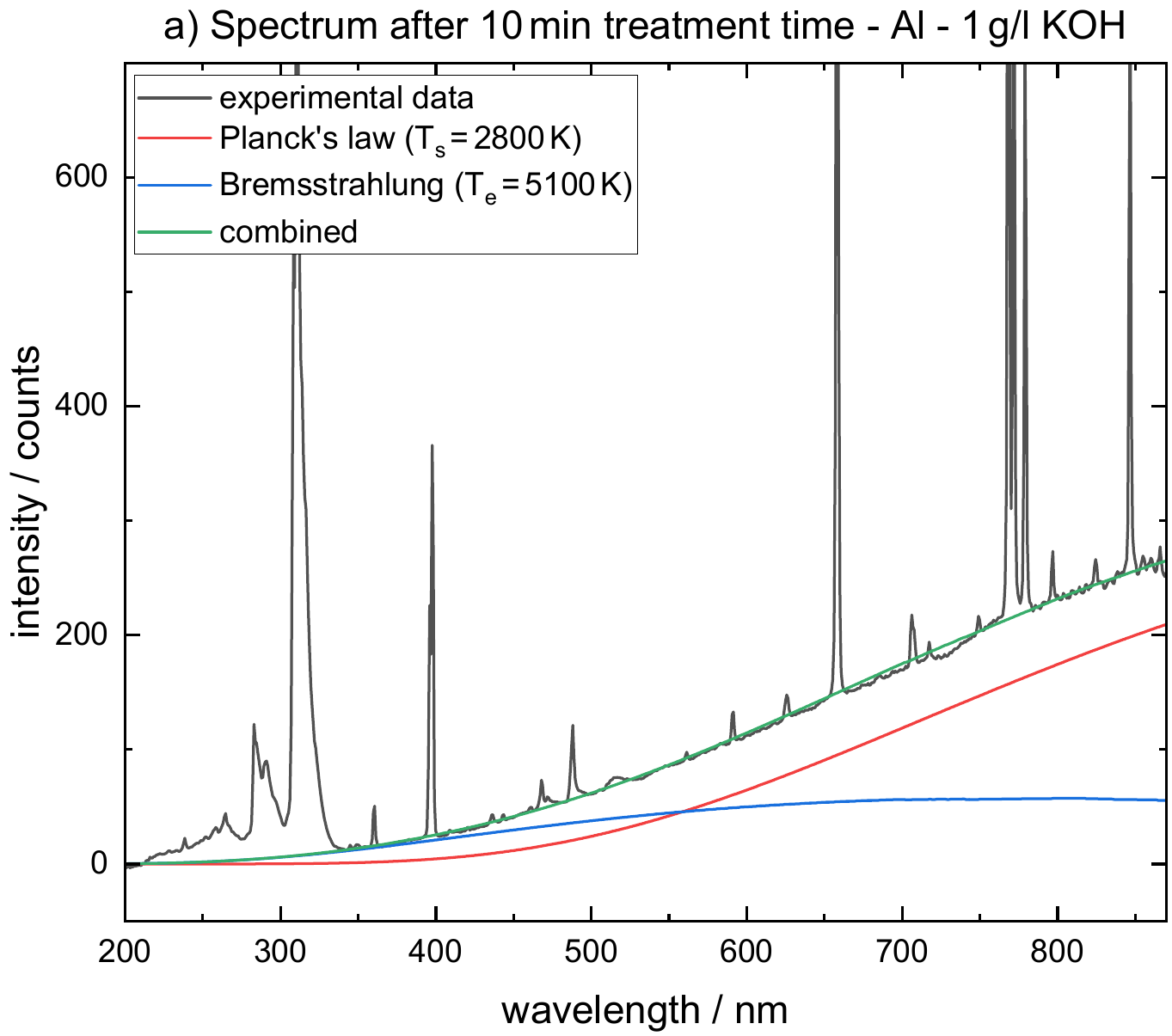}
    \hspace{0.5cm}
    \includegraphics[width=8cm]{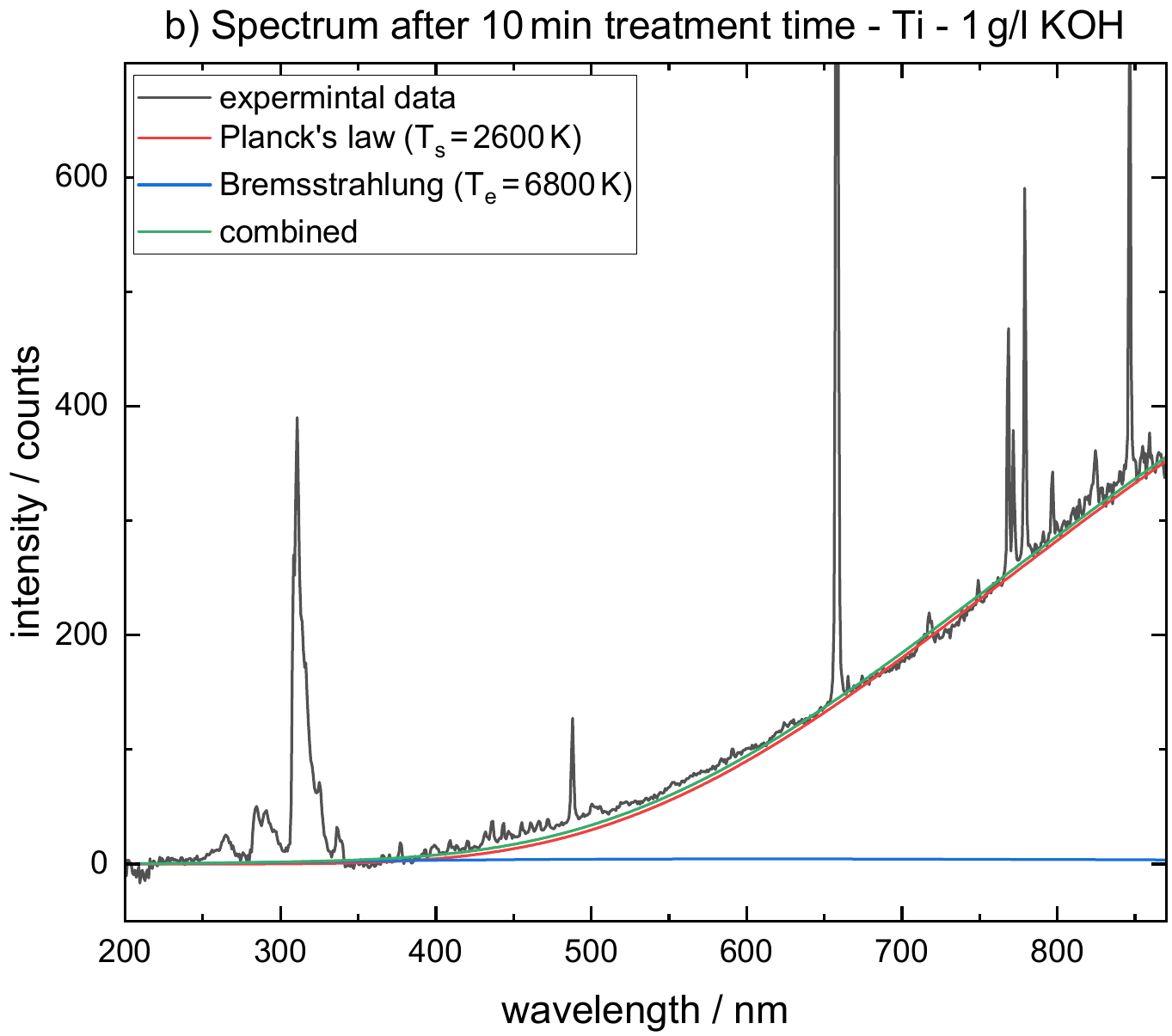}
    \caption{Fitting Bremsstrahlung (blue) and black body radiation (red) to the measured continuum spectrum after 10\,min treatment time with an Al substrate (left) and a Ti substrate (right). Here, the fitting results in an electron temperature of 5100\,K and in a surface temperature of the substrate of 2800\,K for Al, while for a Ti substrate, temperatures of 6800\,K and 2600\,K are achieved, respectively.} 
    \label{sup:cont_Al-Ti}
\end{figure*}

The fitting of Bremsstrahlung and Planck's radiation to the measured continuum spectrum of a PEO process are exemplary shown in figure~\ref{sup:cont_Al-Ti}. The resulting combined lines in green result from an addition of the Bremsstrahlung to Planck's spectrum with the given temperatures.
\section{Morphology of the coating}
Scanning electron microscopy (SEM) was used to analyse the morphology of the coating after the PEO treatment. The SEM images reveal details of the surface structure, including porosity, roughness, and overall uniformity. This supplementary data provides further SEM images, contributing a deeper understanding of the morphology of the coating.
\begin{figure*}
    \centering
    \includegraphics[width=8.2cm]{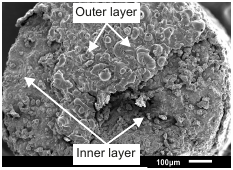}
    \caption{SEM image at 100x magnification for the Al substrate and 2\,g/l\,KOH showing an outer and inner layer of the coating structure. Some coating detachment during sample preparation for the SEM analysis exposes the inner layer and the interface between the inner and outer layer.} 
    \label{sup:SEM:Al_layers}
\end{figure*}
\end{document}


\pagebreak
\widetext
\begin{center}
\textbf{\large Supplemental Materials: Title for main text}
\end{center}
\setcounter{equation}{0}
\setcounter{figure}{0}
\setcounter{table}{0}
\setcounter{page}{1}
\makeatletter
\renewcommand{\theequation}{S\arabic{equation}}
\renewcommand{\thefigure}{S\arabic{figure}}
\renewcommand{\bibnumfmt}[1]{[S#1]}
\renewcommand{\citenumfont}[1]{S#1}

\section{Section 1}
Copy and paste your Supplemental Materials text here \cite{S_RefA}, blah, blah, blah, blah, blah, blah, ...
\begin{equation}
  i\hbar\frac{\partial}{\partial t}\psi(x,t) = -\frac{\hbar^2}{2m}\frac{\partial^2}{\partial x^2}\psi(x,t) + V(x,t) \psi(x,t)
\end{equation}